\documentclass[letterpaper]{JHEP3}
\usepackage{amsmath}
\usepackage{amssymb}
\usepackage{amsfonts}
\usepackage{lscape, graphicx}

\newcommand{\ud}{\textrm{d}}

\newcommand{\commut}[2]{[\,#1\,,\,#2\,]}

\renewcommand{\bar}{\overline}
\newcommand{\tr}{\textrm{tr}}
\newcommand{\Tr}{\textrm{Tr}}

\newcommand{\beqa}{\begin{eqnarray}}
\newcommand{\eeqa}{\end{eqnarray}}
\newcommand{\beq}{\begin{equation}}
\newcommand{\eeq}{\end{equation}}
\newcommand{\vev}[1]{\left\langle\, #1\, \right\rangle}

\title{Lattice chirality, anomaly matching, and more on the (non)decoupling of mirror fermions}
\author
    {
    {
    \def\href#1#2{#2}	% Avoid JHEP bug with footnotes in author list
    Erich Poppitz\footnote{\email{poppitz@physics.utoronto.ca}}~
    and Yanwen Shang\footnote{\email{ywshang@physics.utoronto.ca}}~
           \\Department of Physics, University of Toronto,
    Toronto, ON M5S 1A7, Canada
        }
    }%

\abstract{

\bigskip

We study 't Hooft anomaly matching in lattice models with strong  Yukawa or multi-fermion interactions. Strong non-gauge interactions  among the mirror fermions in a vectorlike lattice gauge theory are  introduced with the aim to obtain, in a strong-coupling symmetric phase, a long-distance unbroken gauge theory with  chiral fermions in  a complex representation. We show how to  use exact lattice chirality  to analyze the  anomaly matching conditions
on chiral symmetry current correlators  at finite lattice spacing and volume.  We perform a  Monte Carlo study 
of the realization of anomaly matching in a toy  two-dimensional model with an anomalous mirror-fermion  content at strong mirror Yukawa coupling. We show that  't Hooft anomaly matching is satisfied, in most of the phase diagram, via the minimal solution in   either the massless fermion or    ``Goldstone" mode, while in some cases  there are extra massless vectorlike mirror fermions.  The mirror spectrum at strong coupling is thus consistent with long-distance unitarity. We discuss the implications of our results  for future studies of the most interesting  case of the decoupling of anomaly-free mirror-fermion sectors.
}

\begin{document}

\maketitle

\section{Introduction, summary, and conclusions}

\subsection{Motivation}

Understanding the dynamics of strongly-coupled chiral gauge theories  is a long-standing problem, where little has been learned since  the 1980s \cite{Raby:1979my}.\footnote{With the exception of  some progress in  chiral  gauge theories with ${\cal{N}}$=$1$  supersymmetry \cite{Shifman:1999mv, Intriligator:2007cp}  and recent advances in the study of semiclassical chiral dynamics \cite{Shifman:2008cx}.}
While not  motivated by a pressing need to explain current data, this issue has its own theoretical interest and appeal. Strong chiral dynamics may gain physical applications in the future, should  results from the LHC indicate that strong interactions play a role in physics near the TeV scale.

This paper is on a topic even more theoretical than that of  studying strong chiral gauge dynamics---the problem of constructing a gauge invariant lattice formulation of chiral gauge theories. Different approaches  and the  status of this subject are reviewed in  \cite{Golterman:2000hr,Luscher:2000hn,Neuberger:2001nb}.

\subsection{The idea}
The idea that will be pursued here is, in its essence, a relatively old one,  
to the best of our knowledge due to Eichten and Preskill (EP) \cite{Eichten:1985ft} (but see also the work of Smit \cite{Smit1}). EP proposed to use the strong-coupling symmetric phases \cite{Fradkin:1978dv, Forster:1980dg,Lang:1981qg,Hasenfratz:1988vc, Stephanov:1990pc},
which occur in lattice theories with non-gauge---e.g., four-fermi, or Yukawa---interactions, to decouple the mirror fermions from a vectorlike gauge theory. The  goal is to thus  obtain,  at long-distances, an unbroken gauge theory with massless chiral fermions in a complex representation.

The strong-coupling symmetric phases  are a lattice artifact, similar to   high-temperature phases in statistical mechanics, where correlations  range  at most over a lattice spacing with no interesting dynamics at long distances.\footnote{Often, the question of the ``triviality" of Yukawa interactions is raised when strong Yukawa interactions are mentioned. We note that  in our  context, triviality---interpreted as the absence of any long-distance physics in the strong-coupling symmetric phase---is  to be advantageously exploited, since the goal of the Yukawa interactions is to keep all mirror fermions at the cutoff scale (see, e.g. \cite{Eichten:1985ft}).}  
The idea of EP, appropriately rephrased to suit our current context,  is to let strong non-gauge  interactions act only on the mirror fermions  and use
 the  short-range   correlations to decouple them from the infrared physics. One can describe the decoupling as  the binding,  due to the strong interactions, of all mirror chiral  fermions into gauge-singlet or vectorlike composites with mass of order the lattice cutoff. In this picture of strong non-gauge mirror dynamics,  due to asymptotic freedom, the gauge interactions play only the spectator role of weakly gauged global symmetries (for a recent study of the decoupling of scalars in a strong-coupling symmetric phase in a simple toy model with gauge fields, see \cite{Poppitz:2008au}). Ideas closely related to that of EP have received attention in the past. In all cases studied, it was found that either the mirrors fail to decouple from the long distance physics or the purportedly light fermions also obtain mass \cite{Golterman:1990zu,Golterman:1991re, Smit2,Smit1993, Golterman:1993th,Golterman:1994at,Golterman:1992yh}.

Despite the failed  early attempts, a proposal to revisit the EP idea in light of the relatively recent progress in  exact lattice chiral symmetry   was made not long ago \cite{Bhattacharya:2006dc} (a suggestion along similar, but not identical, lines had been made   earlier in \cite{Creutz:1996xc}). 
This renewed interest was motivated by the fact that in the pre-1997 formulations of vectorlike gauge theories on the lattice chiral  symmetries were   explicitly broken. The absence of an exact chiral symmetry without doublers  makes it impossible to  split  a lattice Dirac fermion field into chiral components---one to be henceforth called a ``mirror" and the other ``light."  As a result, any strong non-gauge interaction   inevitably  couples to both the ``mirror"  and   ``light" components of the vectorlike fermion. The analysis of the strong dynamics---clearly not purely ``mirror"---of the chiral symmetries and their realization then becomes involved and ambiguous. Most importantly, previous formulations lacked a manifest symmetry explaining why some chiral fermions should stay exactly massless while others could obtain mass; instead, chiral symmetries distinguishing ``light" from ``mirror" modes and protecting the ``light" modes were expected to somehow emerge  at special values of the couplings.

In contrast, the recently discovered exact lattice chiral symmetries without doublers, defined via the Ginsparg-Wilson (GW) relation  \cite{Ginsparg:1981bj} and the Neuberger-Dirac operator \cite{Kaplan:1992bt,Kaplan:1992sg,Narayanan:1993ss,NN2,Neuberger:1997fp}, permit  the splitting of a lattice vectorlike fermion into ``light" and ``mirror" components.   Ward identities for  anomalous \cite{Hasenfratz:1998ri,Fujikawa:1998if,F2,Adams:2000yi,Adams:2001jd} and anomaly-free global symmetries hold  at finite lattice spacing and volume,  ensuring that the chiral symmetries that protect the  light fermions from acquiring mass are exact. Thus, 
  EP-like ideas for the decoupling of the mirror fermions can now be elegantly formulated using exactly chiral lattice  fermions  \cite{Bhattacharya:2006dc}. This elegant realization of symmetries comes at a price, however,  as one still has to show that the strong mirror dynamics   decouples the mirror fermions.  The study of the  mirror theory dynamics is now complicated by the exponential-only\footnote{Exponential locality holds subject to the ``admissibility condition" on the gauge field background \cite{Hernandez:1998et, Neuberger:1999pz}.} locality of the Neuberger-Dirac operator---this is in contrast with local-fermion formulations (whose drawback is the already-mentioned lack of a manifest ``light"-``mirror" split) where the mirror dynamics can be studied via a relatively straightforward strong-coupling expansion.

The lack of an obvious controlled expansion implies that the  mirror fermion dynamics in a formulation of the EP idea with GW fermions has to be investigated via Monte Carlo methods. Due to the high cost of simulations with exactly chiral fermions, such studies are in their infancy. On the other hand, since the issues considered are ones of principle, analyzing two-dimensional models is a sensible first step. Clearly, showing that the ideas work in two dimensions will not prove that four dimensional chiral gauge theories can be similarly formulated, but the results are likely to provide insight into the relevant mirror dynamics.  Another appropriate simplification is to neglect the gauge field fluctuations, since gauge fields play  a spectator role to the strong mirror dynamics. If the strong non-gauge mirror dynamics  gives the mirror fermions mass of order the lattice cutoff (in a manner alluded to above),  asymptotic freedom of the gauge interactions and the exact chiral symmetry of the light  
fermions lead  us to expect that turning on dynamical gauge fields will not significantly affect the mirror dynamics or lift the massless modes. Including dynamical gauge fields would thus only make sense after  the mirror decoupling  in zero gauge background is demonstrated.

\subsection{Its status and a question}\label{status}
The one existing Monte Carlo study, by Giedt and one of us (E.P.) \cite{Giedt:2007qg},  is of the two-dimensional massless Schwinger model with strong  mirror interactions   added, \` a la EP,  in an attempt to  decouple  one of the chiral components of the Dirac fermion. On general grounds, one expects that   it should not be possible to decouple the mirror   and have an unbroken gauge symmetry in this model, since the resulting long-distance theory---the chiral Schwinger model---would be anomalous \cite{Jackiw:1984zi, Halliday:1985tg}. Nevertheless, this potentially ``pathological" setup is the easiest allowing to address questions such as the phase structure of the mirror interactions with GW fermions, because of its simplicity and associated low cost of simulation

In \cite{Giedt:2007qg},  the strong dynamics of the mirror sector of this model was studied via a Monte Carlo simulation,  in a vanishing gauge background. It was found   that the (would-be-gauged) chiral symmetry of the mirror theory is unbroken\footnote{We will liberally use this term in the two-dimensional context here, instead of the more appropriate but unwieldy  ``lack of algebraic order."} and, thus, the strongly-coupled symmetric phase
exists in this model, in a large region of the mirror-coupling space. More intriguingly,  Green's functions of local operators probing the charged fermion spectrum, elementary or composite, showed no evidence for   long-range correlation---and thus for massless charged particles. These results are clearly  puzzling, as gauging the unbroken global chiral symmetry would give rise to an anomalous unbroken gauge theory.

The purpose of this paper is to gain a more detailed  understanding  of the resolution of this puzzle. In particular, we would like to know   what happens when one attempts to decouple mirrors in anomalous representations using non-gauge strong interactions.  There appear to be two possibilities, in a Euclidean rotation-invariant theory: 
{\it a.})~taking the mirror interactions of GW fermions strong leads to an inconsistent theory in the continuum limit\footnote{For example, a nonunitary theory, where the imaginary part of the Euclidean polarization operator is nonlocal, but the real part is local, see Section \ref{matching}.} or {\it b.})~there are massless charged states---Goldstone bosons or fermions---canceling the light fermion anomaly, which were somehow missed in \cite{Giedt:2007qg}.

We first addressed this question  in \cite{Poppitz:2007tu}.  We showed that the partition function of a general vectorlike theory can be split into a light and  mirror part  in an arbitrary gauge background, each part containing a chiral fermion representation. We also demonstrated,   generalizing results of Neuberger \cite{Neuberger:1998xn} and L\" uscher \cite{Luscher:1998pq,Lu2,Luscher:1998du} to arbitrary chiral actions,  that while the  partition function of the vectorlike theory is a smooth function of the gauge background, the 
splitting of the partition function is singular---its light and mirror parts each have singularities in gauge-field space---iff the mirror and light fermion representations are not separately anomaly-free. 
While these results indicated that   anomalous and anomaly-free mirror theories  differ in an essential way,  as Golterman and Shamir pointed out soon thereafter \cite{GS},  the singularity of the light-mirror split  of the partition function does not resolve the puzzle alluded to above (see also the Addendum of \cite{Poppitz:2007tu}). This is because the singularity of the light-mirror split is topological and thus there is freedom---not unlike the ability to move the location of a Dirac string---to render the split of the partition function  locally smooth at any point in gauge field space. Hence, as will be further expanded upon in this paper, there is still  a puzzle in the case where the gauge field is non-dynamical, as in the simulations of \cite{Giedt:2007qg}.

As we will see in this paper, our previous results  \cite{Poppitz:2007tu}  provide us with important tools to further study the attempted  decoupling of an anomalous representation. Since the two-dimensional model studied in \cite{Giedt:2007qg} is the simplest theory with nontrivial mirror dynamics  and is  relatively inexpensive to simulate, understanding the answer to our question in  detail is bound to give us  clues as to the working of the EP proposal formulated with exactly chiral lattice fermions. In particular, we hope to distinguish between the possibilities  ({\it a.}  and {\it b.}) outlined two paragraphs above.

\subsection{Outline and  summary }

We begin  in Section \ref{paradox},  by first describing the deformation of the Schwinger model,  studied in \cite{Giedt:2007qg}, that we call the ``1-0" model. While this model is the subject of our numerical studies here, the analytical results of this paper are more generally applicable. 
In the end of  Section \ref{paradox}, we  review  the main results of \cite{Poppitz:2007tu}, notably the ``splitting theorem" about arbitrary variations of general chiral partition functions with the gauge background, as this technical result is one of the major tools making this study  possible. 
 
The considerations of  the following Section \ref{transversality} are quite general and apply to any EP-like model in either two or four dimensions. 
We first show that  gauge invariance of the full partition function, combined  with the local smoothness of the light-mirror split,  implies certain  anomaly matching  conditions on correlation functions of gauge currents in the mirror theory. In two dimensions, the simplest nontrivial condition  is  on the two-point function in  vanishing gauge background, i.e.~on  the polarization operator.\footnote{In four dimensions with  zero gauge background one would have to consider the three-point function to exhibit the effect of the anomaly; however a study of the polarization operator would still be of use as a probe for the existence of massless charged states.}
The anomaly-matching conditions on  two- or three-point functions of the mirror gauge current  in vanishing gauge background are exact and do not   depend on  the details (i.e.~strength) of the mirror-sector interactions---a simple consequence of the ``splitting theorem" \cite{Poppitz:2007tu}. We note that the anomaly matching conditions  do not require  dynamical gauge fields and introducing infinitesimal gauge backgrounds is sufficient to establish them.

The conditions derived in Section \ref{matching} are the equivalent of 't Hooft anomaly matching  in theories with strong infrared dynamics: whatever the strong mirror dynamics, the mirror spectrum---assuming long-distance unitarity---has to be such that these  anomaly matching conditions hold. A further similarity is that these conditions can be also usefully applied to symmetries that are not  gauged in the target theory.\footnote{In Section \ref{LAST}, we give an example showing how anomaly matching implies that a seemingly viable mirror theory for the anomaly-free ``345" chiral $U(1)$ gauge theory is bound to produce massless mirror modes.}
The fact that conditions like 't Hooft anomaly matching should be obeyed by a strongly-coupled mirror theory on the lattice
should not come as a surprise---that they can be precisely formulated and studied already in finite volume and lattice spacing
is an immediate consequence of the existence of 
  exact chiral lattice symmetries.

 As  usual  in strongly-coupled theories, finding out how  anomaly matching conditions are obeyed is a nontrivial dynamical question, to which we turn next. In  Section \ref{matching}, we review the Goldstone and massless chiral fermion modes of solving anomaly matching and explain how we expect to distinguish between the two in our numerical simulations of the 1-0 model. Then, in Section \ref{generalpiproperties}, we list a number of exact properties of the polarization operators of arbitrary chiral theories (such as our mirror theory), which are independent of the couplings and are derived in  Appendix \ref{polarizationops}.
 The verification of these properties in a Monte Carlo simulation is 
  used to provide important checks on its consistency.

In order to learn how anomaly matching is realized in the 1-0 model,  in Section \ref{nonlocal2} we
begin the study of the mirror polarization operator. We first present the  rather technical  derivation of the mirror-theory gauge two-point function. Since  the details hold for  general theories and may be of interest for future studies, we explain them in several steps in Sections \ref{setup} and \ref{setup1}. The final expression of the mirror polarization operator in terms of correlation functions of the mirror theory (to be computed via Monte Carlo methods) is given in  Section \ref{mirrorPi}. Further details needed to express the correlator via variables  used in the actual simulation are given in  Appendices \ref{notation}, \ref{eigenvectorbasis}, \ref{anomalyderivation}. 

As already stated, our main interest is in the polarization operator of the mirror theory. The real part of the polarization operator at long distances  contains information on the existence and number of massless charged degrees of freedom of the mirror theory. In Euclidean space, the imaginary part   contains  the mirror contribution to  the anomaly. As a check on our  simulation, we have, in each case listed below, verified that the divergence of the imaginary part is exactly as required by  anomaly matching (and is thus equal to the negative of the  divergence of the free light fermion polarization operator).
 
The Monte Carlo  results of this paper are presented in Section \ref{MCresults}.  We begin by first explaining our strategy for looking for massless poles in the polarization operator on an $8 \times 8$ lattice\footnote{As we explain later, the computational demands of the problem limit us to a rather small lattice.} via its small-momentum discontinuity  as a function of direction.  In Fig.~\ref{fig:GWresult}, we show the  small-momentum discontinuity of the real part of the light-theory polarization operator---that of a free chiral GW fermion, defined in eqn.~(\ref{piprime5}) of Appendix A. In later Sections, we compare the analytic free-fermion result of Fig.~\ref{fig:GWresult} to the Monte Carlo results for the strongly interacting mirror sector. 
 
 In Section \ref{MCresultsstrong1}, we show the results of the   Monte Carlo study of the mirror theory in the strong-coupling symmetric phase. This phase---where all mirror-sector global symmetries, apart from the (would-be) gauge symmetry, are explicitly broken by the non-gauge mirror interactions---would be of most interest for mirror-decoupling \` a la EP in anomaly-free models.
We compute the mirror polarization operator and look for the presence of small-momentum discontinuities. The results  for the real part of the mirror polarization operator for different values of the mirror couplings, shown in Figs.~\ref{p_0.1_2}, \ref{p_0.1_5}, \ref{p_0.5_2}, and \ref{p_0.5_5}, are strikingly similar to the free-fermion result of Fig.~\ref{fig:GWresult}. The numerical values and small-momentum discontinuity of the mirror polarization operator indicate that the number of massless charged degrees of freedom in the strong-coupling symmetric phase is the minimal one needed to satisfy anomaly matching. These results  show that the strong mirror dynamics obeys 't Hooft anomaly matching and, in this phase, gives rise to a single massless charged chiral fermion, realizing the possibility {\it b.}) of Section \ref{status} via the minimal solution with massless fermions.
 
 In Section \ref{MCresultsbroken}, we study the mirror theory in the broken phase, where the    ``spin-spin" coupling $\kappa$ is large, see (\ref{Skappa}). The discontinuity of the real part of the polarization operator in Fig.~\ref{p_5_2} is now consistently interpreted as due to a massless scalar (``Goldstone boson") and shows that 't Hooft anomaly matching is obeyed in the Goldstone mode. This is easily explained, as the physics in this phase can be understood in perturbation theory, even at strong Yukawa coupling  \cite{Giedt:2007qg}.

In Section \ref{MCresultsstrong2}, we study the strong-coupling symmetric phase with vanishing Majorana Yukawa  coupling, $h\rightarrow 0$ of eqn.~(\ref{toymodel}). 
The real part of the polarization operator, shown on Fig.~\ref{p_0.5_0.01},  has, as in the $h>1$ phase,  a discontinuity like that of the free chiral fermion. However, the numerical values of the real part  of the polarization operator 
are approximately (we are in a rather small volume) three times larger. This difference is due to the presence of three massless charged chiral degrees of freedom---a massless chiral fermion and a massless vectorlike pair.\footnote{The massless vectorlike pair does not contribute to the anomaly; as a consistency check, we have verified that the divergence of the imaginary part of the polarization operator is as required by anomaly matching.}  
Thus, the spectrum in this phase is consistent with 't Hooft anomaly matching, but has more massless charged fermions than required by anomaly matching. This is explained in Section \ref{anotherview}, where we also show that only at $h=0$ we have analytic control over the  symmetric phase.

The final numerical result of this paper is  presented in Section \ref{MCresultskappa}, where we study the scaling of the real part of the $\kappa$-dependent contributions to the polarization operator deep in the strong-coupling symmetric phase (for decreasing   values of $\kappa\rightarrow 0$). As Fig.~\ref{kappalim} shows,  these terms do not contribute to the discontinuity of the polarization operator. Thus, the discontinuities, and hence the massless fermions, in the strong-coupling symmetric phases arise from the terms containing solely mirror-fermion current-current correlators.

In Section \ref{anotherview}, we  use another representation of the partition function, obtained via a field redefinition. This representation  is particularly useful to analytically interpret the results from Section \ref{MCresultsstrong2}, in the vanishing Majorana coupling strong-coupling symmetric phase (unfortunately,  this is not the phase which would be useful for the study of decoupling in anomaly-free models),  and  explains the origin of the three massless modes found numerically in Fig.~\ref{p_0.5_0.01}.

For convenience of the reader, our conclusions are given in Section \ref{LAST} below. We stress again the main results of this paper---that 't Hooft anomaly matching holds in lattice theories with strong non-gauge dynamics and (assuming unitariry) imposes important constraints on the spectrum at  strong coupling. Our Monte Carlo study of the ``1-0" model mirror dynamics shows that the anomaly matching conditions are obeyed by the mirror theory. In every region of the phase diagram we studied, the strong GW-fermion mirror dynamics  gives rise to massless charged states, realizing option {\it b.}) of Section \ref{status}---in a manner consistent with unitarity of the long-distance theory. Finally, we discuss the implications of these results for anomaly-free models and outline directions for future study.

\subsection{Conclusions, and where do we go next?}
 \label{LAST}

 The discussion of this paper shows that when a fermion formulation with exact lattice chirality is used, the question of decoupling the mirror fermions from a vectorlike gauge theory is intimately intertwined with their contributions to anomalies. While this conclusion is not unexpected, these issues have usually not been central to  past studies (including our own) of attempts to decouple the mirror fermions in strong-coupling symmetric phases. Such studies were usually confined to purely non-gauge strong dynamics at the scale of the lattice spacing, without much emphasis on possible mirror-sector contributions to anomalies. 
This is largely because correlators of global chiral   currents in the mirror theory could not be studied, as the  exact lattice chiral symmetries (which enabled us to  establish many exact properties of these  correlators already at finite lattice spacing and volume) were not known. The main contribution of this paper is the 
precise formulation on the lattice of   't Hooft's anomaly matching conditions, applied to strong mirror interactions, and the study of their solution in a particular example.\footnote{Thus, comparing Fig.~1 to Figs.~2--7 can be considered a Monte Carlo ``proof" of 't Hooft anomaly matching.}
 
This paper shows that models with    mirror Yukawa interactions  formulated via exact lattice chirality are ``smart" enough that, consistent with anomaly matching, in the strong-coupling symmetric phase they lead to massless degrees of freedom, rather than to a nonunitary theory. In our simple toy model, the minimal number of chiral fermions needed to cancel the light fermion anomaly remain massless in the strong-coupling symmetric phase. 
The mirror spectrum emerging from the  cutoff-scale strong dynamics obeys 't Hooft anomaly matching and is consistent with the preservation of (would-be-gauged) unbroken global symmetries.

A final obvious question is what our results here  imply for anomaly-free models. Consider, for example, the ``3-4-5" model, a two dimensional chiral $U(1)$ gauge theory with  $3_{-}$, $4_{-}$ and $5_{+}$ massless fermions (the number indicates the $U(1)$ charge and $\pm$ the chirality). A direct argument predicting massless mirror states, based on the nonvanishing gauge anomaly of the light sector, can not be applied now. What our current knowledge allows us to say for sure is that whether there are 
 massless mirror states---or not---depends  on the implementation of the strong mirror interactions, in particular on their symmetries. 

The  anomaly matching  conditions find another  use in this regard, as we now explain.
Consider implementing the decoupling or mirrors from a vectorlike theory with three Dirac fermions of charges  3, 4, 5 by adding three uncharged mirror fermions and three unitary scalars, i.e.~taking three copies of the 1-0 model and appropriately changing charges/chiralities (one could call the three copies the  3-0, 4-0,  and 5-0 models). From our comments above, it is clear that such an implementation of the 345 model would lead to massless mirror states already in trivial gauge background,  despite two facts that might suggest otherwise:   that the mirror and light spectra are  anomaly-free and that, when the global $U(1)$ appropriate to yield the 345 model is gauged, both the anomaly-free and anomalous global symmetries of such a lattice implementation are as in the target continuum theory. The easiest way to see that massless modes will result at zero gauge background is to  weakly  gauge the three separate global $U(1)$ symmetries present in the 3-0, 4-0, and 5-0 models and demand  consistency of the two-point current correlators, as we did in this paper. 
 
A general way to phrase the condition that   the mirror interactions should obey in order to avoid massless states due to anomaly matching from an extra global symmetry is to demand that when the gauge interactions are turned off, the mirror theory should have no global symmetries other than the global part of the gauge group. With gauge interactions turned off, the implementation of the 345 model of the previous paragraph has two extra global $U(1)$s, which act simultaneously on the light and mirror components and thus impose further conditions on the mirror spectrum implying the existence of  massless modes.

This example  leads us to conjecture  that if the mirror interactions couple the $3_{+}$, $4_{+}$, and $5_{-}$ mirrors  by adding only one scalar and a single neutral $0_{-}$ mirror fermion (needed to have a sensible static limit, see \cite{Bhattacharya:2006dc}), including the most general gauge-invariant couplings breaking all mirror global chiral symmetries, there wouldn't be any massless mirror states in the strong-coupling symmetric phase. At the moment, our strongest argument is that with all the mirror global symmetries explicitly broken,  there is no reason we know of, such as anomaly cancellation of any symmetry, for massless mirrors to  exist for all values of the couplings.  Conversely,  if a future study finds that  massless mirror fermions exist for all values of the mirror couplings, absent any global chiral symmetry in the mirror sector, we expect that there should be  some other reason for this---a symmetry or a deeper dynamical principle we are not yet aware of. Needless to say, it would be of great interest to understand what this principle might be. 
 
Finding out whether our conjecture above is true is left for future work. Two avenues for progress seem promising:~ {\it i.}) coming up with a  theoretical argument why decoupling should always fail and {\it ii.}) verifying or refuting our conjecture via a  numerical ``experiment", similar to the one of this paper, but this time with an anomaly-free model. Since  no  dynamical gauge fields are needed at this stage, the study of (say) the 345-mirror dynamics is quite feasible  given appropriate computer resources and the techniques already developed in \cite{Giedt:2007qg}, \cite{Poppitz:2007tu}, and  the present paper.
\section{The ``1-0" model and the ``splitting theorem"}
\label{paradox}

The Yukawa-Higgs-GW-fermion model  considered here, which we call the ``1-0" model, is a $U(1)$ two-dimensional lattice gauge theory with one Dirac fermion $\psi$ of charge 1 and a neutral spectator  Dirac fermion $\chi$. 
Considering this theory is motivated by its simplicity: it is the minimal Higgs-Yukawa-GW-fermion model in two dimensions which holds the promise to yield, at strong Yukawa coupling, a chiral spectrum of charged fermions and is, at the same time, amenable to numerical simulations not requiring the use of extensive computing resources. 
The fermion part of the action of the ``1-0" model is:
\begin{eqnarray}
\label{toymodel}
S&=& S_{light} + S_{mirror} \\
S_{light} &=&
 - \left( \bar\psi_+ \cdot D_1  \cdot  \psi_+\right) - \left( \bar\chi_- \cdot  D_0  \cdot \chi_-\right) \nonumber  \\
 S_{mirror} &=& - \left( \bar\psi_- \cdot D_1 \cdot  \psi_-\right) - \left( \bar\chi_+ \cdot D_0 \cdot  \chi_+\right)  \nonumber \\
&+& y \left\{ \left( \bar\psi_-  \cdot \phi^*  \cdot \chi_+ \right) + \left( \bar\chi_+ \cdot  \phi \cdot  \psi_- \right)   + h \left[ \left( \psi_-^T \cdot \phi \gamma_2 \cdot  \chi_+ \right) - \left( \bar\chi_+ \cdot  \gamma_2 \cdot  \phi^* \cdot  \bar\psi_-^T \right) \right] \right\} \nonumber~.
\end{eqnarray}
 The chirality components for the charged and neutral fermions are defined by projectors including  the appropriate Neuberger-Dirac operators\footnote{See Appendices \ref{notation}  and \ref{eigenvectorbasis} for our convention.} (charged $D_1$ and neutral $D_0$) for the  barred components, i.e. $ \bar\psi_\pm = \bar\psi (1 \mp \hat{\gamma}_5)/2$ . The brackets denote summation over the lattice sites as well as a spinor inner product.
The field $\phi_x = e^{i \eta_x}$, $|\eta_x| \le \pi$, is a unitary higgs field of unit charge with the usual kinetic term:
\begin{eqnarray}
\label{Skappa}
S_\kappa =  \frac{\kappa}{2}\; \sum_{x} \sum\limits_{\hat{\mu}} \left[ 2 - \left(\; \phi_x^* \; U_{x, x+ \hat\mu} \; \phi_{x+\hat\mu} + {\rm h.c.}\; \right) \right]~.
\end{eqnarray}
The inclusion of both Majorana and Dirac gauge invariant Yukawa terms is due to the requirement that all global symmetries (including those of the mirror fermions) not present in the desired target chiral gauge theory be explicitly broken, see  \cite{Bhattacharya:2006dc}. Moreover, consistent with the symmetries, if the Majorana coupling $h$ vanishes, there are exact mirror-fermion zero modes  for arbitrary backgrounds $\phi_x$, which can not be lifted in the disordered phase  \cite{Giedt:2007qg}. 
The lattice action (\ref{toymodel}) completely defines the theory via a path integral over the charged and neutral fermion fields, the unitary higgs field, as well as the gauge fields. We will not perform the integral over the lattice gauge fields, but will study in detail the variation of the partition function with respect to the gauge background.

 From now on, we will call the fermion fields that participate in the Yukawa interactions the ``mirror" fields---these are the negative chirality component, $\psi_-$, of the charged $\psi$, and the positive chirality component, $\chi_+$, of the neutral $\chi$---while the fields $\psi_+$ and $\chi_-$ will be termed ``light."

Our interest is in the symmetric phase of the unitary higgs theory, expected to occur at $\kappa < \kappa_c \simeq 1$, where the higgs field acts essentially as a random variable (modulo correlations induced by $\kappa \ne 0$ and by  fermion backreaction). Based on experience with strong-Yukawa expansions in theories with naive or Wilson fermions, it is expected  that in the large-$y$, fixed-$h$ limit, there is a symmetric phase. 

The analysis of the 1-0 model of ref.~\cite{Giedt:2007qg} was performed in a vanishing gauge background. The eigenvectors of $\hat\gamma_5$ were explicitly worked out (see also Appendix \ref{eigenvectorbasis}) and used to  manifestly split  the partition function into ``light" and ``mirror."  A Monte Carlo simulation of the mirror sector  at infinite-$y$ and  fixed-$h$  was performed. For small values $\kappa < \kappa_c$ a strong-coupling symmetric phase exists for $h>1$ and for\footnote{Analytic \cite{Gerhold:2007yb} and Monte Carlo \cite{Gerhold:2007gx} studies  of a similar four-dimensional theory with no Majorana coupling, i.e.~at $h=0$, also found evidence for a strong-$y$ symmetric phase.}  $h \rightarrow 0$, while at $h \sim 0.7$ evidence for a Berezinksii-Kosterlitz-Thouless-like transition was found.
The spectrum of the mirror theory was also numerically probed by studying correlators of local operators. The evidence found---using local operators to probe the charged fermion spectrum---pointed toward  decoupling of the mirror sector, with no breaking of the chiral symmetry of the mirror sector (the symmetry to be gauged by the $U(1)$ gauge field). 

These results led to the puzzle already alluded to and to the studies of 
 \cite{Poppitz:2007tu} and the present paper.  In \cite{Poppitz:2007tu}, we studied the splitting of the partition function into ``light" and ``mirror" parts also in an arbitrary nonvanishing  gauge background. 
To remind the reader of our convention (identical to that used in the simulation of  \cite{Giedt:2007qg} but slightly different from the one of \cite{Poppitz:2007tu}), we use the definite-chirality eigenvectors of $\hat\gamma_5$ and the  projectors $\hat{P}_\pm$ on the corresponding spaces:
\beqa\label{uwbasis}
\hat\gamma_5 u_i &=& - u_i ~~,~~ ~~~~~\hat\gamma_5 w_i  = w_i~, \\
\hat{P}_- &=& \sum_i u_i u_i^\dagger ~~,~~ \hat{P}_+  =  \sum_i w_i w_i^\dagger = 1 - \hat{P}_- ~,
\eeqa
where we treat $u,w$ as columns and $u^\dagger, w^\dagger$ as rows. We also use  
the eigenvectors of $\gamma_5$ (the latter are independent of the gauge background) and the associated projectors $P_\pm$:
\beqa
\label{vtbasis}
\gamma_5 v_i &=& v_i ~~,~~~~~~~ \gamma_5 t_i = - t_i~, \\
 {P}_+ &=& \sum\limits_i v_i v_i^\dagger ~~,~~ {P}_-  = \sum\limits_i t_i t_i^\dagger = 1 -  P_+ ~.
\eeqa
Using (\ref{uwbasis}), (\ref{vtbasis}), a general Dirac field $\Psi_x$,  can  be decomposed into chiral components $\alpha^+_i, \alpha^-_i$ via the $\gamma_5$ eigenvectors, while the conjugate spinor field $\bar\Psi_x$ is decomposed into chiral components $\bar{\alpha}^+_i, \bar{\alpha}^-_i$ using the $\hat\gamma_5$ eigenvectors, as follows:
\beqa
\label{psitoc}
\Psi_x &=& \sum_i \alpha^i_+ v_i (x) +  \alpha^i_- t_i(x)~, \\
\bar\Psi_x &=& \sum_i \bar{\alpha}^i_+ u_i^\dagger (x) +  \bar{\alpha}^i_- w_i^\dagger(x)~.
\eeqa
Applying the split (\ref{psitoc}) to the fields $\psi$ and $\chi$ of the ``1-0" model, we note that
only the charged eigenvectors (of both light and mirror fields) depend on the  gauge background. The expansions (\ref{psitoc}) of the ``mirror" fields are explicitly given below:
\beqa
\label{evs1}
\chi_+ &=& \sum_i \beta_+^i v_i  ~, ~~ \bar\chi_+ =  \sum_i \bar\beta_+^i u_i^\dagger[0] ~,\\
\psi_- &= &\sum_i \alpha_-^i t_i~,~~ \bar\psi_-   =  \sum_i \bar\alpha_-^i w_i^\dagger[U] ~.\nonumber 
\eeqa
Clearly, expansions similar to (\ref{evs1}) hold for the ``light" fields as well:
\beqa
\label{evs2}
\chi_- &=& \sum_i \beta_-^i t_i  ~, ~~ \bar\chi_- =  \sum_i \bar\beta_-^i w_i^\dagger[U] ~,\\
\psi_+ &= &\sum_i \alpha_+^i v_i~,~~ \bar\psi_+ =  \sum_i \bar\alpha_+^i u_i^\dagger[U] ~.\nonumber 
\eeqa

After substitution of (\ref{evs1}, \ref{evs2}), the partition function of the model (\ref{toymodel}) splits  as follows:
  \beq
  \label{z01}
  Z[U; y, h] = Z_+[U] \times {1 \over J[U]} \times Z_-[U; y, h]~.
  \eeq
Here $Z_+[U] = \det || (u_i^\dagger[U] \cdot D[U] \cdot v_j) || \times$(similar determinant  for the neutral light spectator $\chi_-$) is the light sector partition function. The jacobian $J$ is a product of  jacobians  for the charged and neutral sectors; see \cite{Poppitz:2007tu} for details. The mirror partition function is denoted by  $Z_-$ and is given, more explicitly, by an integral over the charged mirrors ($\alpha_-, \bar\alpha_-$), neutral mirrors ($\beta_+, \bar\beta_+$), and unitary scalar field:
\beq
\label{mirrorZ1}
Z_-[U;y,h] = \int d^2 \alpha_- \; d^2 \beta_+ \; d \phi \; e^{- S_{mirror} - S_\kappa}~,
\eeq
where the mirror action from (\ref{toymodel}) is expressed in terms of the integration variables $\alpha_-, \beta_+$ and the eigenvectors via (\ref{evs1}).
The mirror fermion integral  is thus a determinant which includes the kinetic term and  Yukawa terms from (\ref{toymodel}) and the mirror partition function is the average of the determinant  over the random (in the disordered $\kappa \rightarrow 0$ phase) unitary field $\phi_x$ (in (\ref{mirrorZ1}), $d \phi$ denotes a path integral over the phases of $\phi$). The mirror partition function (\ref{mirrorZ1}) with $U=1$ was the object used to calculate local mirror observables in the simulations of ref.~\cite{Giedt:2007qg}. 

When $U\ne 1$, the mirror partition function $Z_-$ (\ref{mirrorZ1}) depends on the gauge background through the operators entering $S_{mirror} + S_\kappa$ (the Neuberger-Dirac operator and the associated projectors that appear in (\ref{toymodel})) as well as through the gauge background dependence of the eigenvectors of $\hat\gamma_5$ used to split the partition function ($w_i[A]$, see (\ref{evs1})).
 This dependence was studied in \cite{Poppitz:2007tu}, where an important technical result was derived: under an arbitrary variation of the gauge background, the variation of the mirror partition function, due to the variations of both the eigenvectors and the operators entering the action, factorizes no matter how complicated the mirror partition function is. Explicitly, the ``splitting theorem" states that for an arbitrary variation of the gauge background:
\beq
\label{deltaZchiral}
\delta \log Z_-[U] =  \sum_i (\delta w^\dag_i \cdot  w_i) +  \left< {\delta  S\over \delta O } \; \delta O\right> ~,
\eeq
where ``$\vev{\cdot}$'' denote an expectation value calculated with the partition function $Z_-$ and $O$ collectively denotes the various operators depending on the background which appear in the action (the Neuberger-Dirac operator $D$ and corresponding projectors $\hat{P}_\pm$). This is an important result, as it encodes on the lattice the idea that anomalies do not depend on the action (see the discussion in \cite{Poppitz:2007tu}). Furthermore, as we will see later, the splitting theorem  is indispensible in the calculation of mirror gauge-current correlators in a perturbative expansion in the gauge field.

Coming back to the full 1-0 model partition function, 
we note that because the l.h.s. of (\ref{z01}) is manifestly gauge invariant, so is the r.h.s., since it is obtained  from the l.h.s. simply via a (locally) nonsingular change of variables. 
We know how two of the factors on the r.h.s. transform under gauge transformations---the light partition function  $Z_+[U]$ and the Jacobian $J[U]^{-1}$, for which we have  from \cite{Poppitz:2007tu}, for infinitesimal gauge transformations $\omega$ (as indicated by the $\simeq$ sign):
 \beq
 \label{z02}
 \frac{Z_+[U^\omega]}{J[U^\omega]} \simeq  \frac{Z_+[U]}{J[U]}  {\rm exp}\; \left(  - \frac{i}{2} {\rm Tr} \; \omega \hat\gamma_5 - \sum_i (\delta_\omega w_i^\dagger \cdot  w_i) \right) , 
 \eeq
 where $\delta_\omega u_i $ denotes the variation of $i$-th eigenvector of $\hat\gamma_5$ under a gauge transformation. 
  Therefore, from  (\ref{z02}), and the fact that the l.h.s. of (\ref{z01})  is gauge invariant, it follows that the mirror partition function transforms under gauge transformations as:
  \beq
  \label{z03}
  Z_-[U^\omega; y, h] \simeq Z_-[U; y,h]  \exp\left(   {i \over 2} {\rm Tr} \omega \hat\gamma_5 + \sum_i (\delta_\omega w_i^\dagger \cdot  w_i)  \right)   ,
  \eeq
  independent  not only on the values of the Yukawa couplings ($y, h$) but also most of the details of the mirror action. We also note that the ``splitting theorem" (\ref{deltaZchiral}), valid for general variations,  gives a direct proof of (\ref{z03}) when restricted to gauge transformations \cite{Poppitz:2007tu}.
  
  The gauge variation of the mirror partition function   leads us to the already mentioned paradox.   The exact result (\ref{z03}) shows that the gauge transformation of the mirror partition function is independent of the Yukawa coupling and should precisely cancel that of the light chiral fermion (\ref{z02}). If, at $y\rightarrow \infty$ and $h>1$, the mirror sector only involves heavy degrees of freedom, as the numerical results of \cite{Giedt:2007qg} suggest, 
and if these zero-background results persist for   arbitrarily small gauge backgrounds (as one is inclined to expect based on local smoothness of the basis eigenvectors), then 
the mirror partition function should be  a local functional of the gauge background. But eqn.~(\ref{z03}) argues that  this local functional's gauge variation must precisely cancel the anomaly of the light chiral fermion, which is known to be impossible as the anomaly is not the variation of a local functional.
  
This question was a major motivation for the study of \cite{Poppitz:2007tu}, where we showed that the ``light"--``mirror" split of the partition function is a singular function of the gauge background in  any model where the mirror matter representation is anomalous. In particular, the mirror partition function (\ref{mirrorZ1}), used in the numerical simulation in \cite{Giedt:2007qg}, has a discontinuity precisely at $U=1$. We also conjectured there that the singularity of the mirror partition function might play a role in resolving the paradox, as the results on the mirror spectrum obtained in \cite{Giedt:2007qg} via the singular $Z_-$ (\ref{mirrorZ1}) may not survive an integration over the gauge field or even a perturbative expansion around $U=1$. While this claim may be plausible, 
 if the gauge field is considered as an external background, with small fluctuations around it taken into account perturbatively, a paradox still persists, as is briefly explained in  the Addendum of \cite{Poppitz:2007tu}. This is the issue we want to address here.

\bigskip

\section{Smoothness, the light-mirror split, and anomaly matching}
\label{transversality}

We begin this Section by stating in a more formal way the conflict between the numerical results of \cite{Giedt:2007qg}, which  found no evidence  of long-range correlations in the mirror sector at $y\rightarrow \infty$ and  $h>1$ (when probed with local charged operators), and  eqn.~(\ref{z03}), which states that the mirror partition function of the 1-0 model should effectively  act as a Green-Schwarz term canceling the light sector anomaly. Our goal here is to formulate the anomaly matching conditions in a way useful for further study.

\subsection{Transversality of the full partition function }

The partition function (\ref{z01}) of the 1-0-model is gauge invariant, i.e:
\beq
\label{z1}
\ln Z[A + \delta_\omega A] = \ln Z[A]~,
\eeq
where  $U(x, x+\mu) = e^{i A_\mu (x)}$, with  $\delta_\omega A_\mu(x) = - \nabla_\mu \omega_x$, and $\nabla_\mu \omega_x = \omega_{x+\mu} - \omega_x$. This 
implies that:
\beq
\label{z2}
\sum_{\mu x} \frac{\delta \ln Z[A]}{\delta A_\mu(x)}\;  \nabla_\mu \omega(x) = 0~,
\eeq
where $\nabla^*_\mu \omega_x = \omega_x - \omega_{x-\mu}$.
Taking ${\delta \over \delta \omega(x)}$ of (\ref{z2}) gives:
\beq
\label{z3}
\sum\limits_{\mu} \nabla^*_{\mu x} \; \frac{\delta \ln Z[A]}{\delta A_\mu (x)} = 0~,
\eeq
which, by expanding in $A_\mu$ around $A_\mu = 0$, implies transversality of all $n$-point functions:
\beq
\label{z4}
\sum\limits_\mu \nabla^*_{\mu x} \frac{ \delta^n Z[A]}{\delta A_\mu (x) \delta A_{\mu_1}(x_1) \ldots 
\delta A_{\mu_{n-1}} (x_{n-1})} \bigg\vert_{A=0}= 0~.
\eeq
Eqn.~(\ref{z4}) should apply to the full partition function of the 1-0 model (\ref{toymodel}). The derivation of it assumed that $Z[A]$ is a smooth function of the gauge potential in the vicinity of $A=0$. Smoothness of $Z[A]$ holds because the vectorlike theory has a well defined measure and an action which is smooth with respect to the gauge background.\footnote{We stress again that smoothness holds subject to the ``admissibility condition" on the gauge background \cite{Hernandez:1998et, Neuberger:1999pz}.}  (We note again that the singularity discussed in \cite{Poppitz:2007tu} appears separately in the ``light" and ``mirror" partition functions; the total partition function is that of the anomaly-free vectorlike theory and is  nonsingular.)

\subsection{Local smoothness of the light-mirror split}

The singularity in the ``light"-``mirror" split of the partition funciton  \cite{Poppitz:2007tu} is of topological nature. In the case when the ``light" and ``mirror" sectors are anomalous, there is a topological obstruction to defining their separate fermion measures (or of  the ``measure current" of the $\hat\gamma_5$ eigenvectors) as a globally smooth function of the gauge background.  The mirror partition function and correlators studied in  \cite{Giedt:2007qg} depend on the $\hat\gamma_5$ eigenvectors, which are discontinuous functions of the gauge background when it is turned on.

The topological nature of the singularity means that  its location in the space of gauge backgrounds can be moved around by redefining the phases of the basis vectors (equivalently, of  the measure current). An explicit example is discussed in \cite{Poppitz:2007tu} for the homogeneous
Wilson-line subspace. More generally, due to the smoothness of the $\hat{P}$ projectors, a locally smooth basis of eigenvectors exists at any given point of the space of gauge backgrounds. The possibility of choosing a basis which is  locally  smooth   implies that  in an expansion around $A_\mu = 0$,  (\ref{z4}) can be applied to the split partition function, as we discuss explicitly below. 

In what follows, we present a lengthy derivation of the anomaly matching condition. 
We will use the results obtained here to calculate  the contributions to the polarization operator of the gauge field from the ``light" and ``mirror" fields in later sections. To begin, we  define the polarization operator of the full theory as:
\beq
\label{p1}
\Pi_{\mu \nu}(x,y) \equiv {\delta^2 \ln Z[A] \over \delta A_\mu(x) \delta A_\nu(y) }\bigg\vert_{A=0}~. 
\eeq
Since the full partition function     splits as in eqn.~(\ref{z01}), we have that:
\beq
\label{z6}
\ln Z[A] = \ln Z_+[A] - \ln J[A] + \ln Z_-[A]~.
\eeq
Since the split is locally smooth in the neighborhood of $A_\mu = 0$, eqn.~(\ref{z6}) defines a split of the   polarization operator (\ref{p1}) into ``light" ($\Pi^{+/J}$, with the Jacobian contribution included into the ``light" sector) and ``mirror" ($\Pi^-$) parts:
\beq
\label{z61}
\Pi_{\mu \nu}(x,y) =\Pi_{\mu \nu}^{+/J}(x,y) + \Pi_{\mu \nu}^-(x,y)~.
\eeq

Consider first the ``light" polarization operator.
Under an arbitary infinitesimal change $\delta_\eta$ of the gauge background, the ``light" partition function $Z_+[A] = \det (u_i^\dagger[A] \cdot D[A] \cdot v_j)$ transforms as:
\beqa
\label{z7}
\delta_\eta \ln Z_+[A] &=& {\rm Tr}( P_+ D^{-1} \delta_\eta D) + \sum_j (\delta_\eta u_j^\dagger \cdot u_j) \nonumber \\
&=& {\rm Tr} ( P_+ D^{-1} \delta_\eta D) + j_\eta^u[A]~,
\eeqa 
where $j_\eta^u[A]$ is implicitly defined above.
Similar formulae are derived in \cite{Poppitz:2007tu} (we only give this one in detail due to the different convention of this paper---here we use $\hat{P}_\mp$ to define $\bar\psi_\pm$). We use ``Tr" to denote trace over both spinor and space-time indices, and ``tr" to denote summing over spinor indices only.
We have also introduced the ``measure currents" $j_\eta^{u,w}$:
\beq
\label{measurecurrent}
j_\eta^u[A] \equiv \sum_j (\delta_\eta u_j^\dagger \cdot u_j) ~, ~~ j_\eta^w[A] \equiv \sum_j (\delta_\eta w_j^\dagger \cdot w_j) ~,
\eeq
in terms of which 
the variation of the Jacobian is given as in \cite{Poppitz:2007tu} (modulo the change of convention, see eqn.~(2.16) there):
\beq
\label{z8}
\delta_\eta \ln J[A] =   j_\eta^w[A] + j_\eta^u[A] ~.
\eeq
By combining (\ref{z7}) and (\ref{z8}), we find that the ``light" plus Jacobian contribution to the change of $Z[A]$ of (\ref{z6}) under a gauge transformation (cf. eqn.~(2.17) of \cite{Poppitz:2007tu}) is:
\beq
\label{z9}
\delta_\omega \ln {Z_+[A]\over J[A]} =- j_\omega^w[A] - {i \over 2} \sum_x \omega_x {\rm tr} \hat\gamma_{5 \; xx}[A]~,
\eeq
where $j_\omega$ denotes the measure current (\ref{measurecurrent}) now corresponding to a gauge variation of the background.
Now we take $\delta\over \delta \omega_x$ of (\ref{z9}) to find:
\beq
\label{z10}
{\delta \over \delta \omega_x}\left( \delta_\omega \ln {Z_+[A]\over J[A]} \right)= - {\delta j_\omega^w[A] \over \delta \omega_x} - {i \over 2} {\rm tr} \hat\gamma_{5 xx}[A] ~.
\eeq
Using the identity:
\beq
\label{identity}
{\delta \over \delta \omega_x} \delta_{\omega_x} f[A] = \sum_\mu \nabla^*_{\mu x} {\delta \over \delta A_\mu(x)} f[A], \eeq
eqn.~(\ref{z9}) is clearly the same as:
\beqa
\label{z11}
 \sum\limits_\mu \nabla^*_{\mu x} \; {\delta \ln (Z_+[A] J^{-1}[A]) \over \delta A_\mu (x)} &=& - \sum\limits_\mu \nabla^*_{\mu x} \sum_i ( \delta_{\mu x} w_i^\dagger \cdot w_i)  - {i \over 2} {\rm tr} \hat\gamma_{5 xx}[A] ~,\nonumber \\
 &=& - \sum\limits_\mu \nabla^*_{\mu } j_\mu^w[A]  - {i \over 2} {\rm tr} \hat\gamma_{5 xx}[A] ~, 
\eeqa
where we introduced the shorthand notation:
 \beq
 \label{measurecurrent1}
 \delta_\mu \equiv {\delta   \over \delta A_\mu(x)} 
 \eeq
 for  derivatives $\delta_\mu$ to be used further (note that for brevity we often suppress  the space-time index which we understand to be included in $\mu$). Finally, we expand (\ref{z11}) around $A_\mu = 0$ to linear order, allowed by local smoothness:
\beqa
\label{z12}
\sum\limits_\mu \nabla^*_{\mu x} \; {\delta^2 \ln (Z_+[A] J^{-1}[A]) \over \delta A_\mu (x) \delta A_\nu(y)}\bigg\vert_{A = 0} = - \nabla^*_\mu \;{\delta  j_\mu^w[A] \over   \delta A_\nu(y)}\bigg\vert_{A = 0} - {i \over 2} {\delta {\rm tr} \hat{\gamma}_{5 xx}[A] \over \delta A_\nu (y)}\bigg\vert_{A = 0} ~.
\eeqa
Eqn.~(\ref{z12}), using our definition (\ref{p1}, \ref{z61}) of the polarization operator, is equivalent to:\footnote{We include the Jacobian contribution into the light polarization operator; note that in accordance with the ``splitting theorem" it will cancel with a similar contribution from the polarization operator of the mirror partition function, see \cite{Poppitz:2007tu}.}
\beq
\label{z13}
\sum\limits_\mu \nabla^*_{\mu x} \Pi^{+/J}_{\mu \nu}(x,y) = -  \nabla^*_\mu \; \delta_\nu  j_\mu^w[A]\bigg\vert_{A = 0} - {i \over 2} {\delta_\nu {\rm tr} \; \hat{\gamma}_{5 xx}[A] }\bigg\vert_{A = 0}~,
\eeq
 showing that the ``light" polarization operator is not transverse. 
  
  There are two contributions to the r.h.s. of (\ref{z13}): the first term, proportional to  the derivative of the ``measure current,"   exactly cancels with the identical contributions of the ``mirror" sector---see  eqn.~(\ref{zmirror8})  for the mirror polarization operator.
  The second term on the r.h.s. of (\ref{z13}), proportional to the derivative of the topological lattice field tr$\hat\gamma_{5 xx}$, represents the anomaly of the ``light" fermions.\footnote{While equations similar to (\ref{z13}) will hold  for the non-transverse   higher derivatives of the light partition function, considering only the polarization operator in a trivial gauge background   is sufficient to study the interplay between the anomaly and the light degrees of freedom in 2d. In 4d the trivial gauge background analysis would have to be extended to the three point function in order to capture the effect of the anomaly.}
To make contact with the anomaly in the continuum, we note that the topological lattice field can be expressed  as (the four-dimensional proof of \cite{Luscher:1998kn}
 is trivially downgraded to two dimensions):
\beq
\label{tl1}
{\rm tr}\;  \hat\gamma_{5 xx} = - {1 \over  2 \pi} \epsilon_{\mu\nu} F^{\mu \nu} + \nabla^*_\mu h^\mu[A]~,
\eeq
where $F_{\mu\nu} =\nabla_\mu A_\nu(x) - \nabla_\nu A_\mu(x)$ is the field strength  of the gauge potential $A_\mu(x) = - i \ln U(x,\mu)$,   $h^\mu[A]$ is a gauge invariant local current, and $\epsilon_{12}=1$. An explicit form of $h^\mu[A]$ can be obtained with some work; for example, the part of $h$ linear in $A$ and valid for all momenta, can be derived using the formulae in Appendix \ref{anomalyderivation}, which also presents a derivation of the first term in  (\ref{tl1}).

\subsection{Anomaly matching and  its possible solutions}
\label{matching}

By the local smoothness of the ``light"-``mirror" split and by gauge invariance of the full partition function, the  divergence  of the ``mirror" polarization operator $\Pi_{\mu\nu}^-$ should exactly cancel (\ref{z13}). Since the measure current dependent parts of the ``light" and ``mirror" polarization operators always cancel, the important contribution to the divergence is the one containing (\ref{tl1}) and representing the true effect of the anomaly. We write the cancellation  requirement in the form---taking  the   low-momentum limit and implying a sum over repeated 
indices: 
\beq
\label{piminus1}
i q^\mu \tilde\Pi_{\mu \nu}^-(q) = {1 \over 2 \pi} \epsilon_{\nu \lambda} q^\lambda+~ {\cal{O}}(q^2)~,
 \eeq
 where $\tilde\Pi$ denotes the appropriately defined Fourier transform of the polarization operator---see (\ref{z15}, \ref{divergences}, \ref{tnuk}).
 Now of course the rhs of (\ref{tl1}, \ref{piminus1}) is local, and one wonders if the usual continuum argument that it can not be the divergence of a  local expression applies on the  lattice. A quick argument showing that it does is as follows. Rewrite (\ref{piminus1}) as the set of two equations for the imaginary part of the polarization operator, denoted with the same symbol for brevity,  with $c =-i/(2  \pi)$:
 \beqa
 \label{piminus2}
 \tilde\Pi^-_{22}  &=& - (c  + \tilde\Pi_{12}^-)\; {q_1 \over q_2} ~,\\
 \tilde\Pi^-_{11} &=& (c - \tilde\Pi_{21}^-) \;{q_2 \over q_1} = (c - \tilde\Pi_{12}^-) \;{q_2 \over q_1}, \nonumber
  \eeqa
where we used $\tilde\Pi_{12} =\tilde\Pi_{21}$ due to local smoothness. Locality of $\Pi_{11}$ and $\Pi_{22}$ 
 would then require that $c-\tilde{\Pi}_{12}^- = A q_1 + ...$, $c+ \tilde{\Pi}_{12}^- = B q_2 + ...$, where $A$ and $B$ are arbitrary constants and dots denote higher powers of momenta, 
 leading to:
 \beqa
\label{piminus3}
\tilde\Pi_{12}^- &=& -c + B q_2 + {\cal{O}} (q^2)~,\\
 \tilde\Pi_{12}^- &=& c - A q_1 +{\cal{O}} (q^2)~,\nonumber
\eeqa
conditions  which are clearly incompatible. 

On the other hand, a nonlocal solution of the anomaly conditions (\ref{piminus2}) for the imaginary part of the polarization operator  is given by $\tilde\Pi_{12}^- = c (q_2^2 - q_1^2)/(q_1^2 + q_2^2)$. This is, in fact, the continuum value of the non-transverse part of the polarization operator for an anomalous theory, see eqns.~(\ref{scalar3}, \ref{fermion}) below. 
A  local solution of (\ref{piminus2}) can be found only if $\Pi_{12} \ne \Pi_{21}$, i.e. only if there is a singularity near $A_\mu=0$ such that the second derivatives of $Z_-$ do not commute; this, however, goes against local smoothness of the measure current.  
Thus, our conclusion  is  that local smoothness combined with the nonvanishing anomaly of the light fermion  (\ref{z13})  imply that the imaginary part of the ``mirror" $\Pi_{\mu\nu}^-$ should have a nonlocal part. Eqn.~(\ref{piminus1}), which leads to this conclusion is a mathematical consequence of the gauge invariance of the 1-0 model partition function and the freedom to choose locally smooth $\hat\gamma_5$ eigenvectors. 

We now enumerate the possible solutions of (\ref{piminus1}). First, note that in Euclidean space the anomaly appears in the imaginary part of the polarization operator and this remains true on the lattice, for arbitrary chiral theories formulated with GW fermions (see Appendices \ref{polarizationops}, \ref{anomalyderivation}).  Next, it is well-known from the continuum that in a unitary Lorentz (Euclidean) invariant theory the zero-momentum singularity in the solution of (\ref{piminus1}) is due to either a massless Goldstone boson or a massless fermion \cite{Frishman:1980dq, Coleman:1982yg}. 
The massless scalar or fermion will, of course, also give a nonlocal contribution to the real part of the polarization operator, in addition to the nonlocal contribution to the imaginary part required by the anomaly. On the other hand, if our  complex Euclidean partition function had   no Hamiltonian interpretation, i.e. led to a non-unitary long distance theory, one could  imagine that the imaginary part has a nonlocal contribution while the real part does not. 

Our strategy to look for massless charged mirror modes will be to study both the real and imaginary parts of the polarization operator of the mirror theory. We already know that its imaginary part has a nonlocal contribution  giving rise to  (\ref{piminus1}). The real part of the polarization operator probes the number and nature of the massless charged degrees of freedom and thus gives information of the spectrum. In particular, in two dimensions, massless (Green-Schwarz) scalar or fermion loops lead to $q^2 = 0$ poles in the real part of the polarization operator, see (\ref{scalar3}, \ref{fermion}).
It is this part of 
the mirror polarization operator that will be of most interest to us. 

To recall how this plays out in the continuum and  see what we might expect from our simulations, consider first the example of a    Green-Schwarz scalar theory  with Euclidean partition function:
\beq
\label{scalar1}
Z_{GS}[A] = \int {\cal{D}} \eta \; e^{\int d^2 x (-{\kappa\over 2} (\partial_\mu \eta - A_\mu)^2 + i { \eta\over 2 \pi} F_{12}) }~,
\eeq
such that the partition function is not invariant under gauge transformations. The normalization of the kinetic term is chosen such that it is the naive continuum limit of $S_\kappa$ of (\ref{Skappa}), with $\phi = e^{i \eta}$; note that perturbation theory is good when $\kappa \gg 1$. We define the gauge boson polarization operator as  in (\ref{p1}), and explicitly compute its Fourier transform from (\ref{scalar1}), with the result:
\beq
\label{scalar3}
\tilde\Pi^{\mu\nu}_{GS}\big\vert_{A = 0}(q) =\left(\kappa + {1 \over 4 \pi^2 \kappa}\right)  \left( {q^\mu q^\nu \over q^2} - \delta^{\mu\nu} \right)-  {i \over 2 \pi} \;
{\epsilon^{\nu \rho} q_\rho q^\mu + \epsilon^{\mu \rho} q_\rho q^\nu \over q^2}~.
\eeq 
The polarization operator is not transverse and its divergence is 
$ i q_\mu \tilde\Pi^{\mu\nu}_{GS}\big\vert_{A = 0}(q) = {1\over 2 \pi}  \epsilon^{\nu \rho} q_\rho$, in accordance with (\ref{piminus1}). The nonlocal imaginary part of (\ref{scalar3}) is precisely equal to the solution found in the paragraph below eqn.~(\ref{piminus3}). The real part of (\ref{scalar3}) is the contribution of the massless scalar to the gauge field effective action; notice the shift of the coefficient $\kappa \rightarrow \kappa + {1 \over 4 \pi^2 \kappa}$ due to the anomalous Green-Schwarz term. If our mirror theory has a massless Green-Schwarz scalar, it would have to manifest itself by contributing to both the real and imaginary parts of the mirror polarization operator, as  in (\ref{scalar3}). In fact, we will find (not unexpectedly)  that at large $\kappa \gg 1$, i.e.~in the ``broken" phase of the unitary Higgs mirror theory, this   solution of  anomaly matching is realized.

A massless mirror charged chiral fermion would, similarly,\footnote{The regularization ambiguities  discussed in \cite{Jackiw:1984zi}  do not arise here, since (\ref{fermion}) is the small-momentum limit of the contribution of a  free chiral GW fermion  to the basis-vector independent part of the polarization operator, see Appendices \ref{polarizationops}, \ref{notation}, \ref{anomalyderivation}. It should be clear that we do not claim that (\ref{fermion}) represents the polarization operator of a consistent unitary anomalous theory ((\ref{fermion}) would correspond to the singular $a=1$ case of \cite{Jackiw:1984zi}). 
Instead, eqn.~(\ref{fermion}) is exactly what a massless mirror at $y=0$ would contribute to the full polarization operator of the vectorlike theory. } also  contribute to both the real and imaginary parts of $\Pi_{\mu\nu}$:
\beq
\label{fermion}
\tilde\Pi^{\mu\nu}_{chiral}\big\vert_{A = 0}(q) = {1 \over 2  \pi} \left( {q^\mu q^\nu \over q^2} - \delta^{\mu\nu} \right)-  {i \over 2 \pi} \;
{\epsilon^{\nu \rho} q_\rho q^\mu + \epsilon^{\mu \rho} q_\rho q^\nu \over q^2}~,
\eeq 
where the   real part of $\tilde\Pi_{\mu\nu}(q)$  of the chiral fermion is  equal to one-half that of the Dirac fermion in the Schwinger model.
Note that, as opposed to  the Green-Schwarz scalar (\ref{scalar3}), the coefficients of the real and imaginary parts of the massless chiral fermion $\tilde\Pi_{\mu\nu}$ are  the same---their ratio is $1$ vs. $\simeq 2\pi\kappa$ for the scalar when $\kappa\gg 1$. Again, running ahead, we will find strong evidence that in the disordered phase at $\kappa < \kappa_c$ the mirror spectrum obeys anomaly matching via massless chiral fermions.

In our further study, we will compare our  Monte Carlo results for the mirror Re$\Pi_{\mu\nu}$ with the lattice analogues of eqns.~(\ref{scalar3}) and (\ref{fermion}), which we can easily compute at finite volume and lattice spacing for a free charged unitary scalar and a GW fermion, respectively. We will see that this comparison can already be made on an $8 \times 8$ lattice, providing compelling evidence of what mode the mirror theory chooses to obey the 't Hooft conditions.

\subsection{Exact properties of chiral 
 polarization operators}
\label{generalpiproperties}

We now list some exact properties of polarization operators that hold for general chiral theories, in particular for our mirror theory. These properties are derived in Appendix \ref{polarizationops}.
All equations below refer to the polarization operator in $x$-space and not to their Fourier transforms; recall that we absorb the space-time indices into $\mu$, $\nu$ (see (\ref{measurecurrent1})).

The definition of  $\Pi_{\mu\nu}^- = \delta_\mu \delta_\nu \log Z_-$, see (\ref{p1}, \ref{z6}, \ref{z61}),  implies that $\Pi_{\mu\nu}^-$ is symmetric due to local smoothness of $Z_-$. Furthermore, $\Pi_{\mu\nu}^-$ can be decomposed into a part that is the measure current derivative and the rest, see (\ref{pimirror2}) and (\ref{zmirror8}):
\beq
\label{pgeneral1}
\Pi_{\mu\nu}^- = \delta_\nu j_\mu^w + \Pi_{\mu\nu}^{- \; \prime}~.
\eeq
As shown in Appendix \ref{polarizationops}, $\Pi_{\mu\nu}^{- \; \prime}$ is always a total derivative:
\beq
\label{pgeneral2}
\Pi_{\mu\nu}^{- \; \prime} = \delta_\nu \Pi^{- \; \prime}_\mu~.
\eeq
and, in addition, $\Pi_\mu^{- \; \prime}$ is exactly gauge invariant.
Therefore, proceeding as in the derivation of eqns.~(\ref{z10}, \ref{z11}), we find:
\beq
\label{pgeneral3}
\nabla^*_\nu \Pi_{\mu \nu}^{- \; \prime} = 0~,
\eeq
while with respect to the first index, we have from (\ref{abc}) in Appendix A:
\begin{equation}
\label{abc2}
\nabla^\ast_\mu \Pi^{- \; \prime}_{\mu\nu}=\frac{i}{2}\delta_\nu \tr \hat \gamma^5_{xx}.
\end{equation}
Now, the total mirror $\Pi_{\mu\nu}^-$ is symmetric, but $\delta_\nu j_\mu^w$ and $\Pi_{\mu\nu}^{- \; \prime}$ are separately not, but obey:
\beq
\label{pgeneral4}
  \left( \Pi^{- \; \prime}_ {\mu \nu} -\Pi^{- \; \prime}_ {\nu \mu} \right) = - \delta_\nu j_\mu^w + \delta_\mu j_\nu^w = {\cal{F}}_{\mu \nu}~,
\eeq
where ${\cal{F}}_{\mu \nu}$ is the curvature of the measure current, explicitly given in Appendix \ref{polarizationops}, which is a known local functional of the gauge field whose divergence $\nabla_\mu^* {\cal{F}}_{\mu \nu}$ gives half the anomaly. 
These results imply that the symmetric and antisymmetric parts of $\Pi^{- \; \prime}_{\mu \nu}$ each contribute half of the anomalous divergence (\ref{abc2}) (or (\ref{piminus1})). 

In our numerical simulation we calculate $\Pi_{\mu\nu}^{- \; \prime}$, as the measure-current part is exactly the opposite that of the light theory. 
Since all properties listed in this Section hold independently of the mirror action, in particular of the strength of the mirror couplings, the verification of  (\ref{pgeneral3}), (\ref{abc2}), and (\ref{pgeneral4}), in a numerical simulation at strong mirror couplings provides an important check on its consistency.

\bigskip

\section{Is anomaly matching satisfied? How?}
\label{nonlocal2}

This Section serves two main purposes. First, in Section \ref{setup} we describe the analytical work required to find an expression for the mirror polarization operator in terms of correlation functions of the mirror theory, to be computed via Monte Carlo simulations.  Second, in Section \ref{MCresults},  we present the results of our simulations.

\subsection{Setting up the calculation}
\label{setup}
To appreciate the technical details of the calculation of the mirror polarization operator, we begin by noting that 
if one varies the full theory partition function first, without using a locally smooth light-mirror split, and then substitutes $A_\mu = 0$ and the corresponding $A_\mu = 0$ basis vectors  to  calculate the polarization operator, one finds that the gauge current involves terms that mix ``light" and ``mirror" states. 

This point is already evident in considering the gauge current in the $y=0$ ``light" plus ``mirror" theory, which is given simply by $(\bar\psi \cdot \delta_\mu D \cdot \psi)$. Substituting  the expansions (\ref{psitoc}) of the un-barred spinors  in terms of the $\gamma_5$ eigenvectors $u, t$ ($\psi = \alpha_-^i t_i  + \alpha_+^i v_i$) and  of the barred spinors  in terms of the $\hat\gamma_5$ eigenvectors $w, u$ ($\bar\psi = \bar\alpha_-^i w_i^\dagger + \bar\alpha_+^i u_i^\dagger$), one finds that:
\beqa
\label{crossterms}
(\bar\psi \cdot \delta_\mu D \cdot \psi) &=& \bar\alpha^i_- \alpha^j_- \; (w_i^\dagger \cdot \delta_\mu D \cdot t_j) + \bar\alpha^i_+ \alpha^j_+ \; (u_i^\dagger \cdot \delta_\mu D \cdot v_j) \nonumber \\
&+& \bar\alpha^i_- \alpha^j_+ \; (w_i^\dagger \cdot \delta_\mu D \cdot u_j) + \bar\alpha^i_+  \alpha^j_- \; (u_i^\dagger \cdot \delta_\mu D \cdot t_j) ~,
\eeqa
i.e. the light-mirror  cross terms do not vanish. The $\bar\alpha_\pm \alpha_\mp$ cross terms, upon insertion in a path integral over the fermions $\psi, \bar\psi$,  contribute ``light"-``mirror" loops. This fact would make  the calculation of current-current correlators rather difficult, especially  in the presence of interactions (at $y \ne 0$)  treating different chiralities differently. In particular, if a similar variation was done also in the Yukawa interactions, there would also be ``light"-``mirror" cross terms, making the calculation rather inconvenient, as the ``light" contribution is calculated ``by hand" and the mirror---via Monte Carlo simulations.

Using the locally smooth (in the infinitesimal neighborhood of $A_\mu =0$) basis vectors, together with the ``splitting theorem" of \cite{Poppitz:2007tu}, helps achieve a separation of the  ``light" and ``mirror"  contributions to correlation functions which makes the  calculation manageable. The results for the full partition function do not depend on the basis used, and it is a great convenience to have ``light" and ``mirror"  contributions to current  correlators separated. 
Our goal is to discuss the correlators in the $y=\infty$ mirror theory that  contribute to $\Pi_{\mu\nu}^-$. We use the results of  \cite{Poppitz:2007tu} to simplify the calculation. However, we start with a simple example---the calculation of the mirror polarization operator at $y$=$0$---to illustrate the essential technical points.

\subsubsection{Warm-up: mirror polarization operator  at $y=0$}
\label{setup1}
This warm-up gives a better idea how  the ``splitting theorem" can be used. In the end, of course, adding the ``light" theory piece just reproduces the polarization operator of the massless vectorlike theory at $y=0$ (in a somewhat  ``twisted" left-right separated way, to be useful later).

The $y=0$ ``mirror" partition function---ignoring the neutral mirror, since it will not contribute to the polarization operator when $y=0$---is: 
\beqa
\label{zmirror1}
Z_- &=& \int d \alpha_- d \bar\alpha_- e^{- S}~,\\
S &=&  - \bar\alpha_-^i \alpha_-^j (w_i^\dagger \cdot \hat{P}_+ \cdot D \cdot t_j) \nonumber
\eeqa
and its first variation wrt $A_\mu (x)$ (below, we use $d^2 \alpha_-$ to denote $\prod_i d \alpha_-^i d \bar\alpha_-^i$) is:
\beqa
\label{zmirror2}
\delta_\mu Z_- &=&  \int d^2 \alpha_-\;  \bar\alpha_-^i \alpha_-^j (\delta_\mu w^\dagger_i \cdot \hat{P}_+ \cdot D \cdot t_j) e^{-S}\\
&+&\int d^2 \alpha_- \; \bar\alpha_-^i \alpha_-^j (w^\dagger_i \cdot \hat{P}_+ \cdot
 \delta_\mu   D \cdot t_j) e^{-S} \nonumber \\
 &+& \int d^2 \alpha_- \; \bar\alpha_-^i \alpha_-^j (w^\dagger_i \cdot \delta_\mu \hat{P}_+ 
  \cdot D \cdot t_j) e^{-S} \nonumber~. 
\eeqa
The first line above equals $Z_- \sum_i (\delta_\mu w^\dagger_i \cdot w_i) = Z_- J_\mu^w$ by the splitting theorem (\ref{deltaZchiral}).

We note that the second line in (\ref{zmirror2}) represents  a different chiral partition function (as, by the criterion of  \cite{Poppitz:2007tu},  it is a partition function whose variation wrt $w^\dagger$ is orthogonal to $u^\dagger$) and the splitting theorem can be used while calculating the second variation of this term. Finally, the third line  in (\ref{zmirror2}) vanishes, because $D \cdot t_j = D \cdot P_- t_j = \hat{P}_+ \cdot D \cdot t_j$, combined with $\delta_\mu \hat\gamma_5 \cdot \hat\gamma_5 = - \hat\gamma_5 \cdot \delta_\mu \hat\gamma_5$, implies that after  $\hat{P}_+$ is pushed through $\delta_\mu \hat{P}_+$ it becomes $\hat{P}_-$ and annihilates  $w^\dagger$.  We then have from the above comments:
\beqa
\label{zmirror3}
\delta_\mu Z_- &=&  Z_- j_\mu^w \\
&+& \int d^2 \alpha_-\; \bar\alpha_-^i \alpha_-^j \; (w^\dagger_i \cdot \hat{P}_+ \cdot
 \delta_\mu D \cdot t_j) e^{-S}\nonumber~.
\eeqa

Next, we calculate the second variation. A very important technical point  is that the splitting theorem can be applied iff the variation of the partition function due to the change of the basis vector is orthogonal to the opposite chirality basis vector. This is the reason we keep the factor of $\hat{P}_+$ in the second line in (\ref{zmirror3}) even though it might appear as a tautology here; the price to pay is that we have to account for the projector's variation when we calculate the second variation of $Z_-$, as we will see below. Thus we find:
\beqa
\label{zmirror4}
\delta_\mu  \delta_\nu Z_- &=& \delta_\nu Z_- \times j_\mu^w \nonumber \\
 &+& Z_- \times \delta_\nu j_\mu^w  \nonumber \\
 &+& J_\nu^w \times \int d^2 \alpha \; \bar\alpha^i_- \alpha^j_-  (w^\dagger_i   \cdot \delta_\mu D \cdot t_j) e^{-S} \\
&+&\int d^2 \alpha  \; \bar\alpha^i_-  \alpha^j_-  (w^\dagger_i \cdot \delta_\mu D \cdot t_j)  \; \bar\alpha^p_-  \alpha^q_-  (w^\dagger_p  \cdot \delta_\nu D \cdot t_q) \;e^{-S} \nonumber \\
&+& \int d^2 \alpha  \; \bar\alpha^i_-  \alpha^j_-  (w^\dagger_i \cdot \delta_\nu \hat{P}_+ \cdot \delta_\mu D \cdot t_j) e^{-S} \nonumber \\
&+&  \int d^2 \alpha \; \bar\alpha^i_-  \alpha^j_-  (w^\dagger_i \cdot \delta_\nu \delta_\mu D \cdot t_j) e^{-S} \nonumber ~. 
 \eeqa
 We used the splitting theorem on the third line above, wrt the chiral partition function given by the second line in (\ref{zmirror3}). Armed with (\ref{zmirror3}) and (\ref{zmirror4}) we can now compute
 the $y=0$ ``mirror" theory polarization operator, defined by (\ref{p1}) with $Z \rightarrow Z_-$:
 \beqa
 \label{pimirror1}
\Pi_{\nu\mu}^-&=& \delta_\nu j_\mu^w    \nonumber \\
&+& \langle \bar\alpha_-^i \alpha_-^j \rangle \left((w^\dagger_i \cdot \delta_\nu \delta_\mu D \cdot t_j) + (w^\dagger_i \cdot \delta_\nu \hat{P}_+ \cdot  \delta_\mu D \cdot t_j) \right) \\
&+&\left(  \langle \bar\alpha_-^i \alpha_-^j \bar\alpha_-^p \alpha_-^q \rangle  - \langle \bar\alpha_-^i \alpha_-^j\rangle  \langle  \bar\alpha^p_- \alpha_-^q\rangle \right) (w^\dagger_i \cdot   \delta_\mu D \cdot t_j)   (w^\dagger_p \cdot  \delta_\nu D \cdot t_q)  ~, \nonumber \eeqa 
 where brackets ``$\vev{\cdot}$'' denote expectation values calculated with the mirror partition function. It is immediately seen (by 
recalling \cite{Poppitz:2007tu}), using
 $\langle \bar\alpha^i \alpha^j\rangle =|| (w^\dagger_k \cdot D \cdot t_l)^{-1}||_{ij} =
 (t^\dagger_i \cdot D^{-1} \cdot w_j)$, that the polarization operator of the mirror can be written as:
 \beqa
 \label{pimirror2}
 \Pi_{\mu\nu}^- &=& \delta_\nu j_\mu^w + {\rm Tr} (\delta_\mu \delta_\nu D \cdot P_- \cdot D^{-1} \cdot \hat{P}_+ )+ {\rm Tr} (\delta_\nu \hat{P}_+ \cdot \delta_\mu D \cdot P_- \cdot D^{-1} \cdot \hat{P}_+)  \\
 &-& {\rm Tr}( \delta_\mu D \cdot P_- \cdot D^{-1} \cdot \hat{P}_+ \cdot \delta_\nu D \cdot P_-\cdot D^{-1} \cdot \hat{P}_+ ) \nonumber~.
 \eeqa
 Many of the projectors above can be dropped due to the Ginsparg-Wilson relation, but were left in for comparison with (\ref{pimirror1}). 
 
 Now, it is straightforward to compute $\Pi_{\mu\nu}^-$ directly from the expression for the $y=0$ charged-mirror partition function $Z_- = \det || (w_i^\dagger \cdot D \cdot t_j) ||$ and see complete agreement with (\ref{pimirror2}); we leave this as an exercise for the reader.
 The calculation in this section was done in such detail in order to  see  the agreement with the much faster direct calculation alluded to above and to emphasize that the term with the variation of the projector in (\ref{pimirror2}) is crucial for this agreement---this projector was inserted  to allow use of the splitting theorem when performing the second variation of the mirror partition function. 
 
Of course, for the case of interest $y\ne 0$, we do not have an expression for    $Z_-$ as simple as $ \det ||(w_i^\dagger \cdot D \cdot t_j) ||$, so the best we can hope for is to cast the mirror $\Pi_{\mu\nu}^-$ into a form similar to (\ref{pimirror1}), which will give the mirror polarization operator in terms of  mirror theory correlation functions. These  can be computed via a Monte Carlo simulation at zero gauge background, by first computing explicitly the various functions of position and momentum appearing in (\ref{pimirror1}), such as $(w^\dagger_i \cdot \delta_\nu \hat{P}_+ \cdot  \delta_\mu D \cdot t_j)$, and then using the existing code of \cite{Giedt:2007qg}.

\subsubsection{Mirror polarization operator at $y\ne 0$}
\label{mirrorPi}
The  1-0 model action is  given in (\ref{toymodel}).  Substituting (\ref{evs1}) into the mirror action, we find the terms in the mirror action that depend  on the gauge field:
\beqa
\label{mirroractiongauge}
S_{mirror} &=& -\; \bar\alpha_-^i \alpha_-^j \; (w^\dagger_i[A] \cdot D[A] \cdot t_j) + y \; \bar\alpha_-^i \beta_+^j \; (w_i^\dagger[A] \cdot \hat{P}_+[A] \cdot \phi^* v_j) \nonumber \\
&&-\; y h \; \bar\beta^j_+ \bar\alpha^i_- \; (u_j^\dagger \gamma_2 \cdot \phi^* \cdot \hat{P}_+^T[A] \cdot w_i^*[A]) + S_\kappa[A] \\
&&+\;    {\rm A_\mu-independent} \;, \nonumber
\eeqa
where $S_\kappa$ is the kinetic action for $\phi$. The mirror partition function is defined in (\ref{mirrorZ1}).
For the first variation of $\ln Z_-[A]$ we  find (brackets $\vev{\cdot}$ now denote expectation values with the full mirror partition function  (\ref{mirrorZ1})): 
\beqa
\label{zmirror7}
\delta_\mu \ln Z_- &=&\; ~~j_\mu^w \nonumber \\ 
&&+\; \langle\; \bar\alpha_-^i \alpha_-^j \; (w_i^\dagger \cdot \hat{P}_+ \cdot \delta_\mu D\cdot t_j) \rangle \nonumber \\
&&+ \;{\kappa \over 2} \; \langle \; (\phi^* \cdot \delta_\mu U \cdot \phi) +{\rm  h.c.}\; \rangle \\
&&-\; y \;  \langle \; \bar\alpha_-^i \beta_+^j \; (w_i^\dagger \cdot \hat{P}_+ \cdot \delta_\mu \hat{P}_+ \cdot \phi^* v_j)\rangle \nonumber \\
&&- \;y h \; \langle \bar\alpha^i_- \bar\beta_+^j \; (u^\dagger_j \gamma_2 \cdot \phi^* \cdot \delta_\mu \hat{P}_+^T \cdot \hat{P}_+^T \cdot w_i^*) \nonumber~,
\eeqa
where the first two lines on the r.h.s. appear just as in  (\ref{zmirror3}).
Now we have to perform a second variation, carefully use the splitting theorem several times, and finally compute $\Pi_{\mu\nu}^- = \delta_\nu \delta_\mu \ln Z_-[A]\big\vert_{A=0}$, watching for the many cancellations. This calculation  straightforwardly follows the already established rules and after some tedious algebra we obtain the mirror polarization operator:
\beqa
\label{zmirror8}
\Pi_{\mu\nu}^- &=& \delta_\nu j_\mu^w \nonumber \\
&+& \langle \bar\alpha_-^i \alpha_-^j \rangle \; (w_i^\dagger \cdot (\delta_\mu \delta_\nu D + \delta_\nu \hat{P}_+ \delta_\mu D)\cdot t_j) \nonumber \\
&+&  \langle \bar\alpha_-^i \alpha_-^j  \bar\alpha_-^k \alpha_-^l \rangle^C \; (w_i^\dagger \cdot \delta_\mu D \cdot t_j) (w_k^\dagger \cdot \delta_\nu D \cdot t_l) \nonumber \\
&+&\frac{\kappa}{2} \; \langle ( \phi^* \cdot \delta_\nu \delta_\mu U \cdot \phi) + {\rm h.c.} \rangle \nonumber \\
&+&\frac{\kappa^2}{4} \; \langle \left[ (\phi^* \cdot  \delta_\mu U \cdot \phi) + { \rm h.c.}\right] \left[  ( \phi^* \cdot  \delta_\nu U \cdot \phi) + { \rm h.c.}\right] \rangle^C \nonumber \\
&+&\frac{\kappa}{2}  \left[ \langle  \bar\alpha_-^i \alpha_-^j ( \left( \phi^* \cdot  \delta_\nu U \cdot \phi) + { \rm h.c.}\right) \rangle^C (w_i^\dagger \cdot  \delta_\mu  D \cdot t_j) + (\mu \leftrightarrow \nu) \right] \nonumber \\
&-& y \; \langle \bar\alpha_-^i \beta^j_+\; (w_i^\dagger \cdot \delta_\nu (\hat{P}_+ \delta_\mu \hat{P}_+) \cdot \phi^* v_j)\rangle  \\
&-& y h \; \langle \bar\alpha_-^i \bar\beta_+^j \; (u_j^\dagger \gamma_2 \cdot \phi^* \cdot\delta_\nu (\delta_\mu \hat{P}^T_+ \cdot \hat{P}^T_+ ) \cdot  w_i^*)\rangle\nonumber \\
&-&y \left[ (w_i^\dagger \cdot \delta_\mu D \cdot t_j)  \langle \bar\alpha^i_- \alpha^j_- \bar\alpha^k_- \left( \beta_+^l (w_k^\dagger \cdot \delta_\nu \hat{P}_+ \cdot \phi^* v_l) + h \;  \bar\beta_+^l (u_l^\dagger \gamma_2 \cdot \phi^* \cdot \delta_\nu \hat{P}_+^T \cdot w_k^*)  \rangle^C  \right)  \right. \nonumber \\
& & \left. ~ ~~ + ( \mu \leftrightarrow \nu) \right] \nonumber \\
&+& y^2 \; \langle \left( \bar\alpha^i_- \beta_+^j (w_i^\dagger \cdot \delta_\mu \hat{P}_+ \cdot \phi^* v_j) + h \bar\alpha_-^i \bar\beta_+^j (u_j^\dagger \gamma_2 \cdot \phi^* \cdot \delta_\mu \hat{P}_+^T \cdot w_i^*) \right) \times   \nonumber \\
&&~~   ~~~ \left( \bar\alpha^k_- \beta_+^l (w_k^\dagger \cdot \delta_\nu \hat{P}_+ \cdot \phi^* v_l)+ h \bar\alpha_-^k \bar\beta_+^l (u_l^\dagger \gamma_2 \cdot \phi^* \cdot \delta_\nu \hat{P}_+^T \cdot w_k^*) \right)\rangle^C \nonumber \\
&-& \frac{y \kappa}{2} \left[ \langle \left[ (\phi^* \cdot  \delta_\mu U \cdot \phi) + { \rm h.c.}\right] \left[\bar\alpha_-^i \beta^j_+ (w_i^\dagger \cdot \delta_\nu \hat{P}_+ \cdot \phi^* v_j) + h \bar\alpha_-^i \bar\beta_+^j (u_j^\dagger \gamma_2 \cdot \phi^* \cdot \delta_\nu \hat{P}_+^T \cdot w_i^*) \right] \rangle^C \right. \nonumber \\
&&  \left. ~ ~~ + (\mu \leftrightarrow \nu) \right] \nonumber ~,
\eeqa
with $\langle ... \rangle^C$ indicating the connected part.
The first three lines are identical to (\ref{pimirror1}).

The  form of the mirror polarization operator (\ref{zmirror8}) simplifies somewhat in the $y=\infty$ limit of our simulation when many 
terms vanish. At infinite Yukawa, the mirror theory conserves the number of 
$+$ fermions minus the number of $-$ fermions, implying that all correlators with an unequal number of $+$ and $-$ fermions vanish as $y \rightarrow \infty$.
Therefore, we have a simpler expression for the mirror polarization operator (\ref{zmirror8}), which we write as a sum of several terms:
\beq
\label{zmirror9}
\Pi_{\mu\nu}^-\big\vert_{y\rightarrow\infty} = \delta_\nu j_\mu^w + \Pi_{\mu\nu}^\prime ~, ~~~\Pi_{\mu\nu}^\prime \equiv \Pi_{\mu\nu}^y   + \Pi_{\mu\nu}^{\kappa y} + \Pi_{\mu\nu}^\kappa~,
\eeq
where the fermion current-fermion current contribution $\Pi_{\mu\nu}^y$ is:
\beqa
\label{zmirror12}
\Pi_{\mu\nu}^y =
&-& y \; \langle \bar\alpha_-^i \beta^j_+\; (w_i^\dagger \cdot \delta_\nu (\hat{P}_+ \delta_\mu \hat{P}_+) \cdot \phi^* v_j)\rangle - y h \; \langle \bar\alpha_-^i \bar\beta_+^j \; (u_j^\dagger \gamma_2 \cdot \phi^* \cdot\delta_\nu (\delta_\mu \hat{P}^T_+ \cdot \hat{P}^T_+ ) \cdot  w_i^*)\rangle\nonumber \\
&+& y^2 \; \langle \left( \bar\alpha^i_- \beta_+^j (w_i^\dagger \cdot \delta_\mu \hat{P}_+ \cdot \phi^* v_j) + h \bar\alpha_-^i \bar\beta_+^j (u_j^\dagger \gamma_2 \cdot \phi^* \cdot \delta_\mu \hat{P}_+^T \cdot w_i^*) \right) \times   \nonumber\\
&&   ~~~~ \left( \bar\alpha^k_- \beta_+^l (w_k^\dagger \cdot \delta_\nu \hat{P}_+ \cdot \phi^* v_l) + h \bar\alpha_-^k \bar\beta_+^l (u_l^\dagger \gamma_2 \cdot \phi^* \cdot \delta_\nu \hat{P}_+^T \cdot w_k^*) \right)\rangle^C ~ ,
\eeqa
the scalar current-scalar current contribution $\Pi_{\mu\nu}^\kappa$ is:
\beqa
\label{zmirror13}
\Pi_{\mu\nu}^\kappa = && \frac{\kappa}{2} \; \langle ( \phi^* \cdot \delta_\nu \delta_\mu U \cdot \phi) + { \rm h.c.} \rangle \nonumber \\
 &+& \frac{\kappa^2}{4} \; \langle \left[ (\phi^* \cdot  \delta_\mu U \cdot \phi) + { \rm h.c.}\right] \left[  ( \phi^* \cdot  \delta_\nu U \cdot \phi) + { \rm h.c.}\right] \rangle^C ~, 
\eeqa
and, finally, the mixed fermion  current-scalar current contribution $\Pi_{\mu\nu}^{y \kappa}$ is:
\beqa
\label{zmirror14}
&&\Pi_{\mu\nu}^{y \kappa} = \nonumber \\
 &&- \frac{y \kappa}{2} \left\{ \langle \left[ (\phi^* \cdot  \delta_\mu U \cdot \phi) + { \rm h.c.} \right]  \left[\bar\alpha_-^i \beta^j_+ (w_i^\dagger \cdot \delta_\nu \hat{P}_+ \cdot \phi^* v_j) + h \bar\alpha_-^i \bar\beta_+^j (u_j^\dagger \gamma_2 \cdot \phi^* \cdot \delta_\nu \hat{P}_+^T \cdot w_i^*) \right] \rangle^C \right. \nonumber \\
&&+  \left.  (\mu \leftrightarrow \nu)\right\} ~.
\eeqa
It should be clear that many factors (i.e.~basis vectors 
$w^\dagger$, $u^\dagger$, $v$ and operators $\hat{P}_+$) can be taken out of the correlators as they do not depend on the integration variables ($\alpha_-$, $\bar\alpha_-$, $\beta_+$, $\bar\beta_+$, $\phi$) but are simply functions of momenta; we do not explicitly do this here for brevity.

A few comments are now in order. The measure current contribution exactly cancels  the similar contribution of the light sector, the first term in (\ref{z13}), and, as stated many times, all the ``action" is  in the other terms. Consider first the terms independent on $\kappa$. 
The correlators that enter $\Pi_{\mu\nu}^y$  (\ref{zmirror12}) scale as $1/y$ and $1/y^2$, respectively, so  their contributions to $\Pi_{\mu\nu}^-$ are $y$-independent. As for the $h$-dependence  of the large-$y$ limit of $\Pi_{\mu\nu}^-$, we only   know for sure that $h$ should drop out of the divergence of the polarization operator (it is difficult to see how this can be proved starting from the explicit expression of the mirror polarization operator, but it must be true, as a consequence of the gauge invariance of the full ``light"+``mirror" partition function and the local smoothness of the split). 
Finally, we expect that the terms proportional to $\kappa$,  $\Pi_{\mu\nu}^\kappa$ and $ \Pi_{\mu\nu}^{y \kappa}$ vanish in the $\kappa \rightarrow 0$ limit: it is not clear how a $1/\kappa$ singularity in the scalar Green's functions could come about if there is indeed a disordered phase in this limit (as per the results of \cite{Giedt:2007qg}); we numerically test the $\kappa \rightarrow 0$ behavior of $\Pi_{\mu\nu}^\kappa+\Pi_{\mu\nu}^{y \kappa}$ in Section \ref{MCresultskappa} and confirm this expectation.

In our numerical simulation, we compute $\Pi_{\mu\nu}^\prime$ of eqn.~(\ref{zmirror9})  
and look for non-local terms in its real part. 
Before stating the results of the simulations in the next Section, we note that in order to compute the scalar and fermion correlators in (\ref{zmirror9}), one needs to choose a representation of $\gamma$-matrices, a basis of $\hat{\gamma}_5$-eigenvectors,  work out the perturbative expansion of the Neuberger-Dirac operator to second order in the gauge field, and finally, express the correlators (\ref{zmirror12}--\ref{zmirror14}) in terms of these quantities. We give our notation and conventions in   Appendices \ref{notation} and \ref{eigenvectorbasis}  in sufficient detail to complete this straightforward but tedious calculation.

\subsection{Monte Carlo results for the mirror polarization operator at strong coupling}
\label{MCresults}

\begin{figure}
\includegraphics[width=6in]{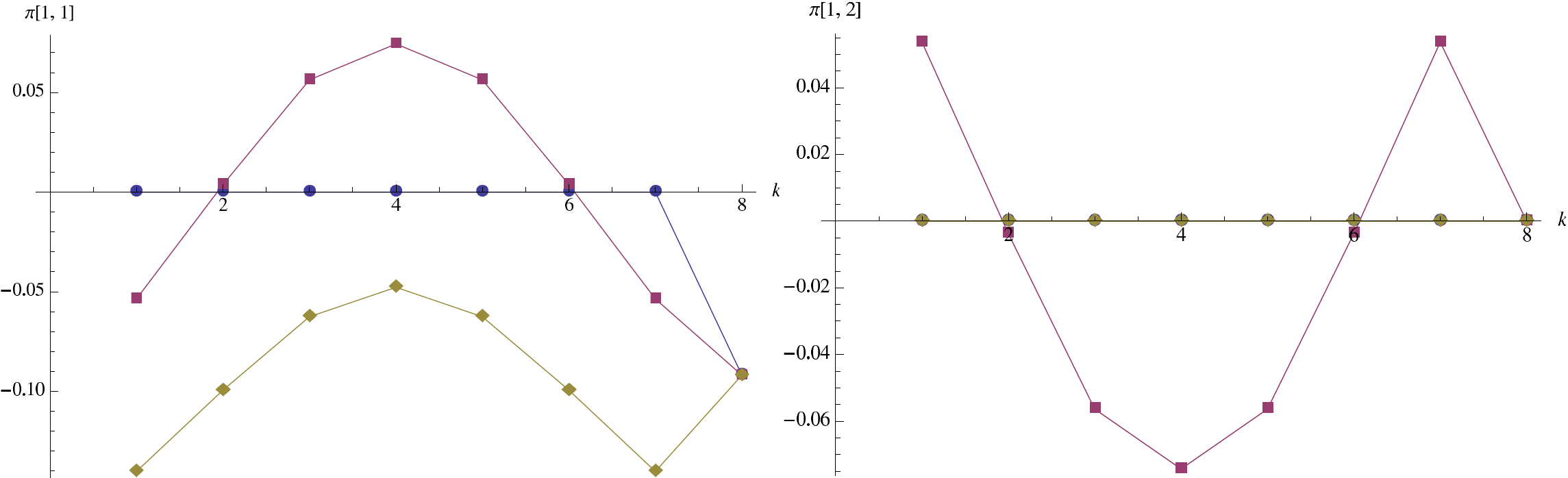}
\caption
{
The real parts of   $\tilde\Pi_{11}$ (left plot) and $\tilde\Pi_{12}$ (right plot), the Fourier components of the real part of the basis-vector independent part of polarization operator  of a free chiral GW fermion on an $8^2$ lattice, as a function of momentum. The $k$-dependence is plotted along three lines  in momentum space---the circles and horizontal squares correspond to lines approaching the origin at $0^0$ ($k_2 = 0$) and $45^0$ ($k_1 = k_2 =k$) wrt $x$-axis, respectively, while the tilted squares are taken on a line with $k_1 = 0$ (at $90^0$). Clearly, the discontinuity at small $k$ (i.e., at $k= 7$ and $k=1$) is as predicted by the continuum expression---for example, the value of $2 C$ read off $ \tilde\Pi_{11}(90^0)$ agrees well with the continuum expression. We note that in these plots (generated by Mathematica), momenta are labeled by $1...8$ instead of $0...7$ as in the plots presenting the Monte-Carlo results; an identification $0=8$ should be made for comparison.
}
 \label{fig:GWresult} 
\end{figure}
 
In the continuum,  the contribution to the Fourier transform of the real part of the polarization operator due to massless particles is:
\beqa
\label{picont1}
\tilde\Pi_{\mu\nu}(k) = 2 C \;   {\delta_{\mu\nu} k^2 - k_\mu k_\nu\over k^2}~, ~~2 C_{GS\; scalar} \simeq - {\kappa}, ~~ 2C_{ch.ferm.} \simeq-{1 \over 2 \pi} \simeq -0.159, 
\eeqa 
where the constant $C$ depends on the number of massless degrees of freedom and the normalization is chosen for later convenience. The values of $C$ given are for a  single Green-Schwarz scalar and a single chiral fermion, inferred by comparison with (\ref{fermion}) and (\ref{scalar3}).
The polarization operator (\ref{picont1})    has a directional singularity as $k \rightarrow 0$:
\beqa
\label{picont2}
\tilde\Pi_{11}(\phi)\big\vert_{k \rightarrow 0} &=&  C (1 - \cos 2 \phi)~ \nonumber \\
\tilde\Pi_{21}(\phi)\big\vert_{k \rightarrow 0} &=&  -C  \sin 2 \phi ~, 
\eeqa
where $\phi$ is the angle of approach to the origin  measured from the positive-$k_1$ axis. 
Thus, from (\ref{picont2}) we expect that on the lattice, if there are massless particles, the following relations will hold as $k \rightarrow 0$:
\beqa
\label{picont3}
\tilde\Pi_{11}(45^0) &=& - \tilde\Pi_{21}(45^0) = C \; , \nonumber \\
\tilde\Pi_{11}(90^0) &=& 2 C \; , \\
\tilde\Pi_{11}(\; 0^0) &=& \tilde\Pi_{21}(0^0) = \tilde\Pi_{21}(90^0) = 0 \; .\nonumber
\eeqa

To test the relations (\ref{picont3}) on the lattice, we begin by first computing the real part  of the polarization operator of free massless GW fermions, which is simply equal to $1/2$ the Dirac GW fermion polarization operator and the analytical expression for which is given in  (\ref{piprime5}) of  Appendix \ref{polarizationops}. 
On Fig.~\ref{fig:GWresult}, we plot the real part of the Fourier transform of the real part\footnote{This cumbersome expression is,  unfortunately, unavoidable: real and imaginary parts of $\Pi_{\mu\nu}^\prime$ are defined in $x$-space, but then the finite lattice Fourier transform (\ref{FT}) of   Re$\Pi_{\mu\nu}^\prime$ has both a real and imaginary part and becomes real only in the continuum limit.} of $\Pi_{\mu\nu}^\prime$ for a free chiral GW fermion (\ref{piprime5}); the Fourier transform is precisely defined in \ref{FT}. 
In order to make the calculation of (\ref{piprime5}) well-defined, we imposed antiperiodic boundary conditions on the fermions, which results in a well-defined value at $k=0$, but does not remove the directional singularity as $k \rightarrow 0$. 

It is clear from Fig.~\ref{fig:GWresult} that  the polarization operator approaches different limits as one approaches the origin at different angles.
The discontinuity at small momenta is clearly visible already on 
the rather small lattice used. The results agree well with the expectation of eqn.~(\ref{picont3}).
 The value of $C$ inferred from  $k=1$ on an $8^2$ lattice does approximately match the one of the continuum result for a chiral fermion, given in (\ref{picont1}); a rather precise numerical agreement can be seen on somewhat larger lattices (e.g., virtually identical on $32^2$), however, we only plot the $8^2$ result since this is the lattice size of our numerical simulation.

If one performs a similar calculation for a massive fermion, the $k \rightarrow 0$ discontinuity   disappears, in accord with the  expectation that a massive-particle loop contributes $\Pi_{\mu\nu} \sim m^{-2} (\delta_{\mu\nu}k^2 - k_\mu k_\nu)$.
 
In what follows, we will compare the results for $\tilde\Pi_{11}$ and $\tilde\Pi_{21}$ for the free chiral GW fermion with the results of the Monte Carlo simulation of the same components 
of the mirror polarization operator. While in the text  we refer to $11$ and $12$ components of the polarization operator, in the Figures  and the captions, the replacement $\tilde\Pi_{11} \rightarrow \tilde\Pi_{00}$, $\tilde\Pi_{21} \rightarrow \tilde\Pi_{10}$, will be used.

\subsubsection{A few words about the simulation}

All our simulations are on an $8\times 8 $ lattice. The reason we only consider such a small lattice is that the computation of $\Pi_{\mu\nu}$ is rather demanding, because of the large number of momentum sums that occur in the correlators, most notably in the fermion-fermion current-current correlator of eqn.~(\ref{zmirror9}). Nevertheless, when combined with the analytic results of the previous Sections and the Appendices, even this rather small lattice is sufficient for a qualitative study of the mirror spectrum and in particular, an identification of the massless mirror states.

The code used to generate configurations was developed by J. Giedt for the study of \cite{Giedt:2007qg} and uses the cluster algorithm to generate $XY$-model configurations with a reweighting of the fermion determinant.  
The fermion measure is taken into account through
determinant reweighting:
\beqa
\langle{{\cal{O}}}\rangle = \frac{ \langle{{\cal{O}} \det M}\rangle_\eta }{\langle{ \det M}\rangle_\eta}.
\eeqa
Here, $\cal{O}$ is any observable, and $\langle{ \cdots }\rangle_\eta$
denotes an expectation value with respect to 
the measure of the XY model (cf.~eqn.~(\ref{Skappa}) with $U \equiv 1$), 
\beqa
d \mu(\eta) = Z_{XY}^{-1} \left(\prod_x d\eta_x\right) \exp (-S_\kappa).
\eeqa
We monitor the reliability
of this method in several ways.
 We measure the
autocorrelation time for reweighted quantities
$\langle {\cal{O}} \det M\rangle_\eta$, as
well as $\langle{\det M}\rangle_\eta$, to be certain that the configurations
remain independent with respect to the new measure.  We perform a jackknife error analysis of
the averages $\langle{{\cal{O}}}\rangle$ that are obtained, gathering sufficient
data to keep errors small.

An additional package was developed by one of us (Y.S.) so that the original code was adapted to work on the Sunnyvale computer cluster at CITA, where the momentum sums over different configurations were performed with hundreds of processors in parallel.  In order to reduce the statistical errors, 16000 independent configurations were generated and used in the calculation of the averages.

\subsubsection{Strong-coupling symmetric phase ($h>1, \kappa \ll \kappa_c$): massless chiral fermion  }
\label{MCresultsstrong1}

\begin{figure}
\includegraphics[width=6in]{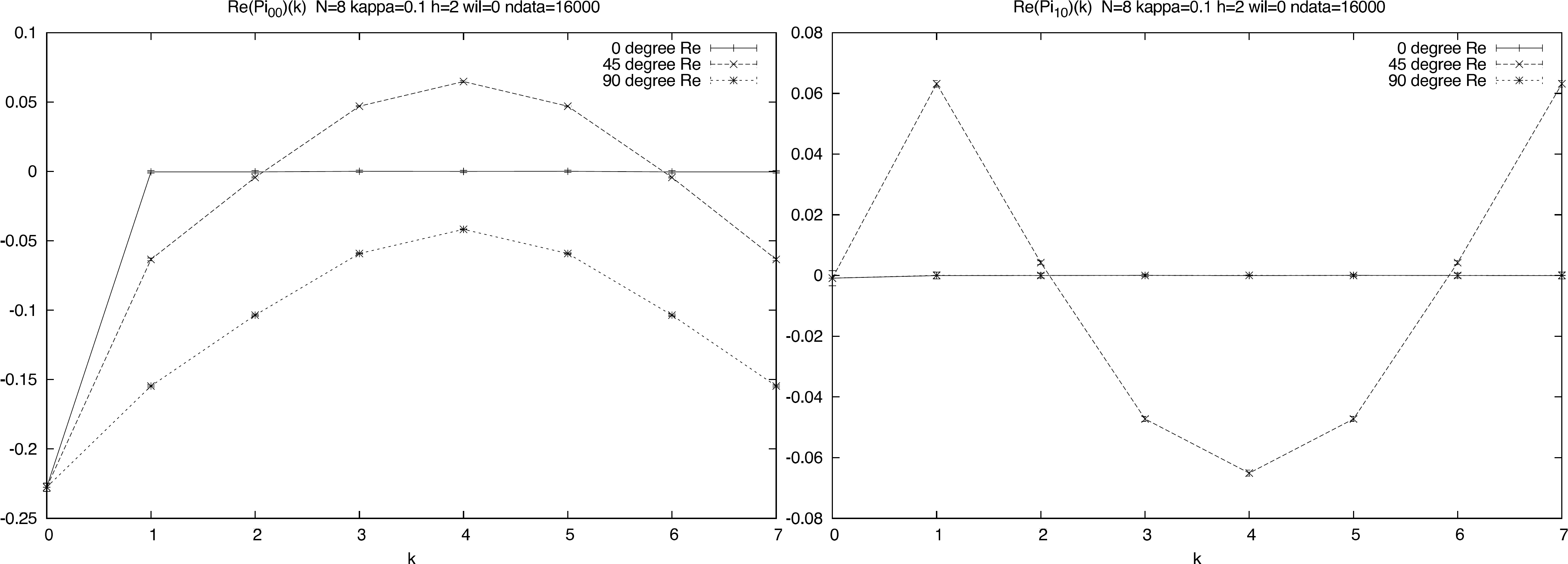}
\caption
{The real parts of $\Pi_{00}$ and $\Pi_{10}$ of the mirror for $\kappa = 0.1$, $h=2$, as a function of momentum approaching the origin in  different directions: symmetric phase, massless fermion (compare with Fig.~1.).
}
 \label{p_0.1_2} 
\end{figure}
\begin{figure}
\includegraphics[width=6in]{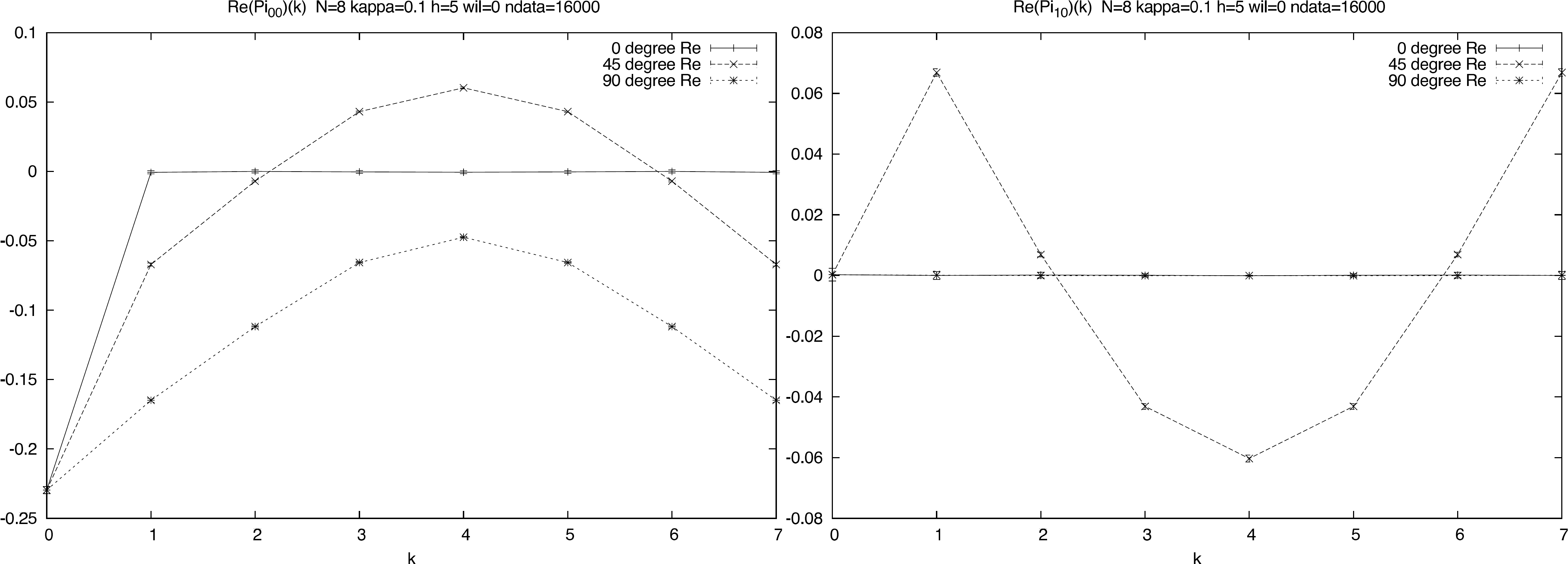}
\caption
{The real parts of $\Pi_{00}$ and $\Pi_{10}$ of the mirror for $\kappa = 0.1$, $h=5$:   symmetric phase, massless fermion (compare with Fig.~1.).
}
 \label{p_0.1_5} 
\end{figure}
\begin{figure}
\includegraphics[width=6in]{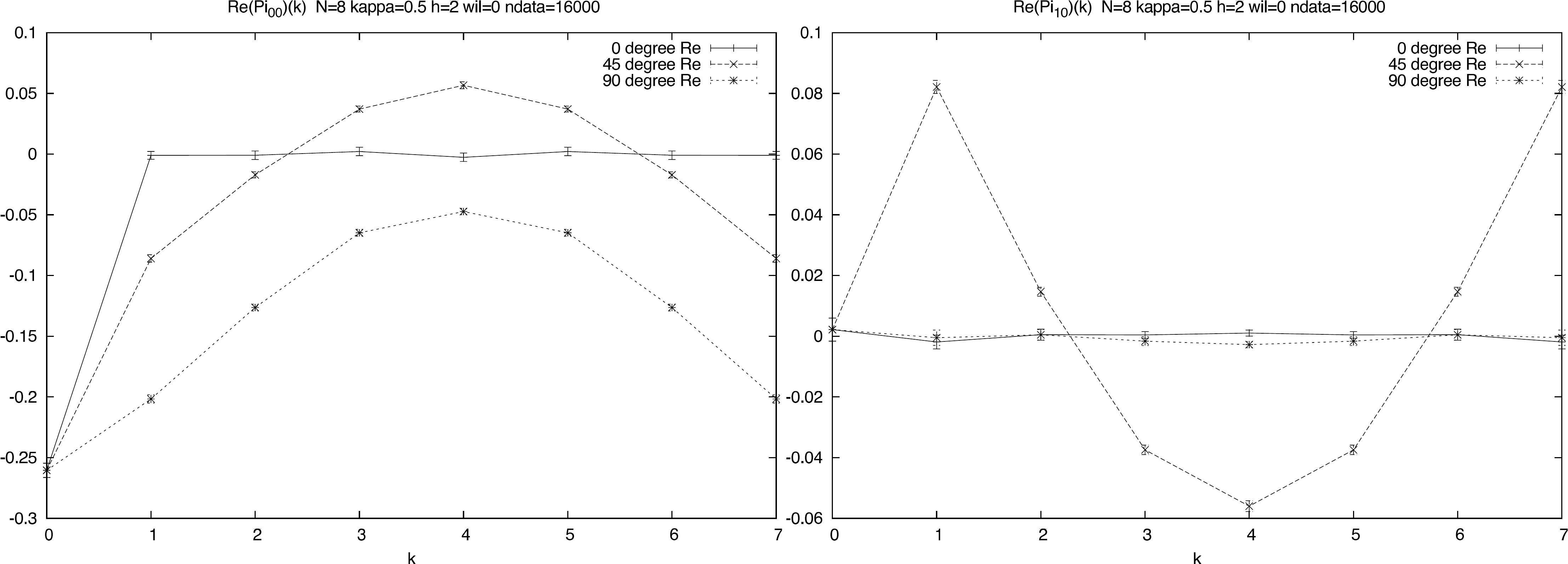}
\caption
{The real parts of $\Pi_{00}$ and $\Pi_{10}$ of the mirror for $\kappa = 0.5$, $h=2$: symmetric phase, massless fermion (compare with Fig.~1.).
}
 \label{p_0.5_2} 
\end{figure}
\begin{figure}
\includegraphics[width=6in]{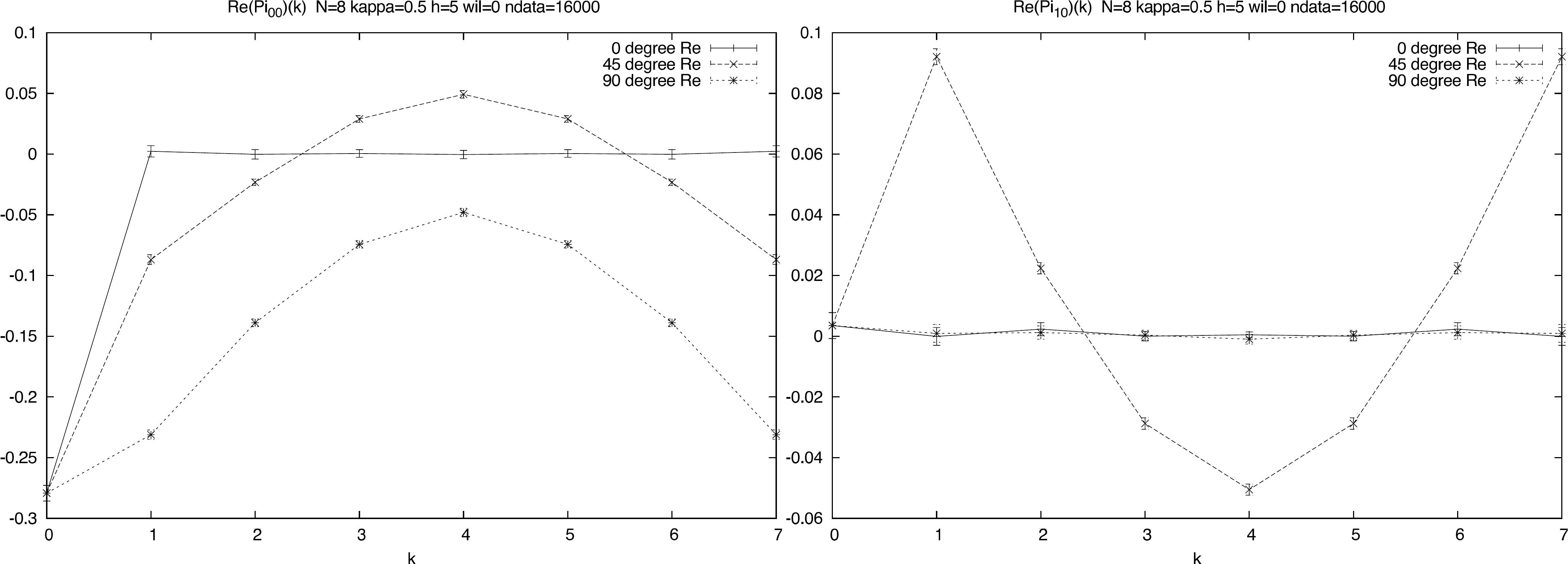}
\caption
{The real parts of $\Pi_{00}$ and $\Pi_{10}$ of the mirror for $\kappa = 0.5$, $h=5$: symmetric phase, massless fermion (compare with Fig.~1.).
}
 \label{p_0.5_5} 
\end{figure}

We now compare Fig.~\ref{fig:GWresult} to the numerical simulation of $\Pi_{\mu\nu}^\prime$ of the mirror theory, an expression for which in terms of mirror correlation functions is given in  (\ref{zmirror9}). On Figs.~\ref{p_0.1_2} and \ref{p_0.1_5} we show the same components of the polarization operator as on Fig.~\ref{fig:GWresult} for $\kappa = 0.1$ and $h=2, 5$, respectively. On Figs.~\ref{p_0.5_2} and \ref{p_0.5_5} we show the results for $\kappa = 0.5$ and $h=2, 5$, respectively. In all cases, measurements of the susceptibilities   \cite{Giedt:2007qg} indicated that the theory is in the strong coupling symmetric phase.  

Thus, we observe that for $\kappa < 1$ and $h>1$, i.e. in the strong-coupling symmetric phase found in \cite{Giedt:2007qg}, the polarization operator of the mirror theory is qualitatively and quantitatively close to the one in the free GW-theory of Fig.~\ref{fig:GWresult}. The discontinuity at small-$k$ is as expected from the continuum formula for the real part of a chiral fermion polarization operator and the actual numerical values of Re$\Pi_{\mu\nu}$ are also close to the ones for the free fermion (for example, the values of $2 C$ inferred from $ \tilde\Pi_{11}(90^0)$, see (\ref{picont3}), are closer to the continuum value for $\kappa = 0.1$, deeper in the symmetric phase, than for $\kappa = 0.5$). The real part shows slight variations with $h$, as the plots for $h=2$ and $h=5$ show, as well as with $\kappa$ (when $\kappa<1$), as the plots for $\kappa = 0.5$ and $\kappa  = 0.1$ show. In each case, the ratio of $\tilde\Pi_{11}(90^0)/\tilde\Pi_{11}(45^0) \simeq 2.5$, are in agreement with  the ratio for free GW fermions of Fig.~\ref{fig:GWresult}  and somewhat larger than the continuum ratio of 2.

In every case, the divergence of the imaginary part,  as well as eqns.~(\ref{pgeneral3}), and (\ref{pgeneral4}), were numerically checked to be obeyed by $\Pi_{\mu\nu}^\prime$ up to the errors of the simulation. We conclude that in this phase the mirror theory satisfies 't Hooft anomaly matching in the massless chiral fermion mode.

\subsubsection{``Broken" phase ($h>1$, $\kappa \gg \kappa_c$): Green-Schwarz scalar}
\label{MCresultsbroken}
\begin{figure}
\includegraphics[width=6in]{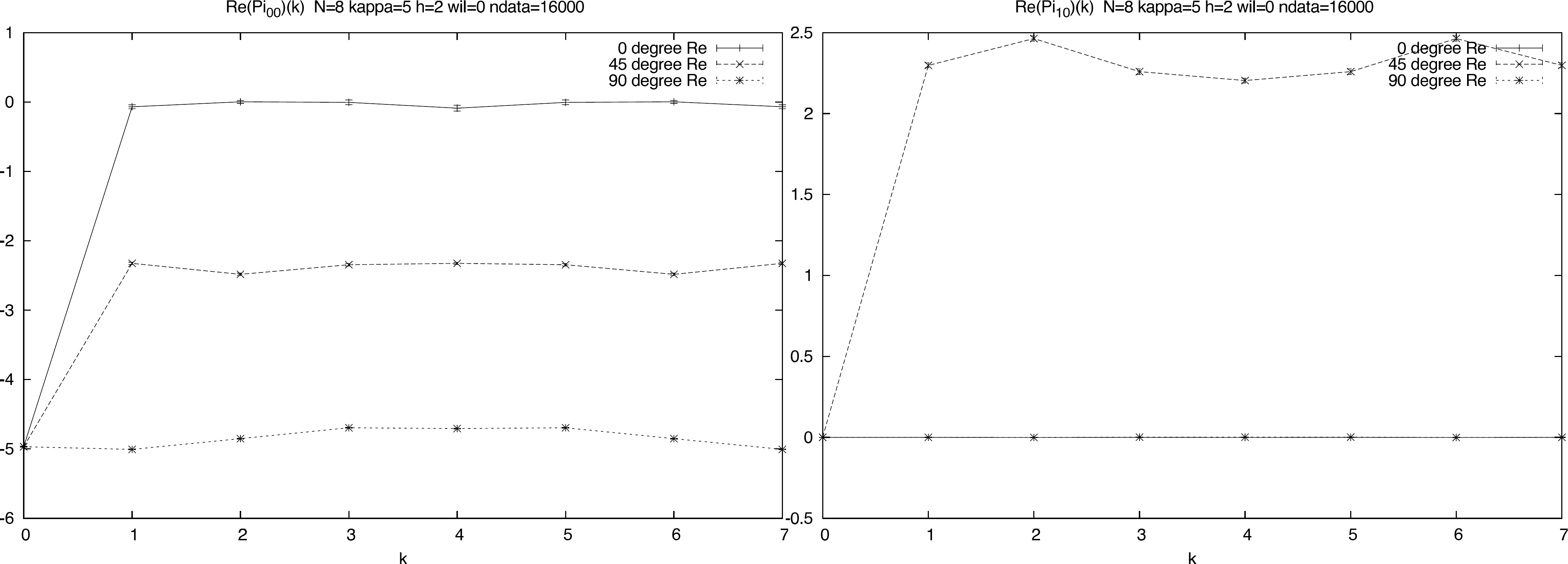}
\caption
{The real parts of $\Pi_{00}$ and $\Pi_{10}$ of the mirror for $\kappa = 5$, $h=2$: broken phase, massless scalar.
}
 \label{p_5_2} 
\end{figure}
The results for the mirror polarization operator in the broken phase are shown on Fig.~\ref{p_5_2}, for $\kappa=5, h=2$.
Taking $\kappa \gg 1$ takes the unitary Higgs field into an algebraically ordered  (``broken") phase. The studies of \cite{Giedt:2007qg} showed that the results of simulations in this phase agree very well with perturbation theory (a $1\over \kappa$ expansion). This continues to be the case here---note that the values of $\Pi_{\mu\nu}$ at small momenta scale with $\kappa$, in contrast to the symmetric $\kappa < 1$ phases, see eqn.~(\ref{scalar3}). From $\tilde\Pi_{11}(90^0)$ we infer   $2 C \simeq - 5$, in good agreement with the value for $\kappa =5$ of eqn.~(\ref{picont1}).
Thus, we conclude that  the discontinuity of $\Pi_{\mu\nu}^\prime$ on Fig.~\ref{p_5_2} in the ``broken" phase is due to a massless scalar.

\subsubsection{Strong-coupling symmetric phase ($h \rightarrow 0$, $\kappa \ll \kappa_c$): massless  chiral fermion and massless vectorlike  pair
}
\label{MCresultsstrong2}
\begin{figure}
\includegraphics[width=6in]{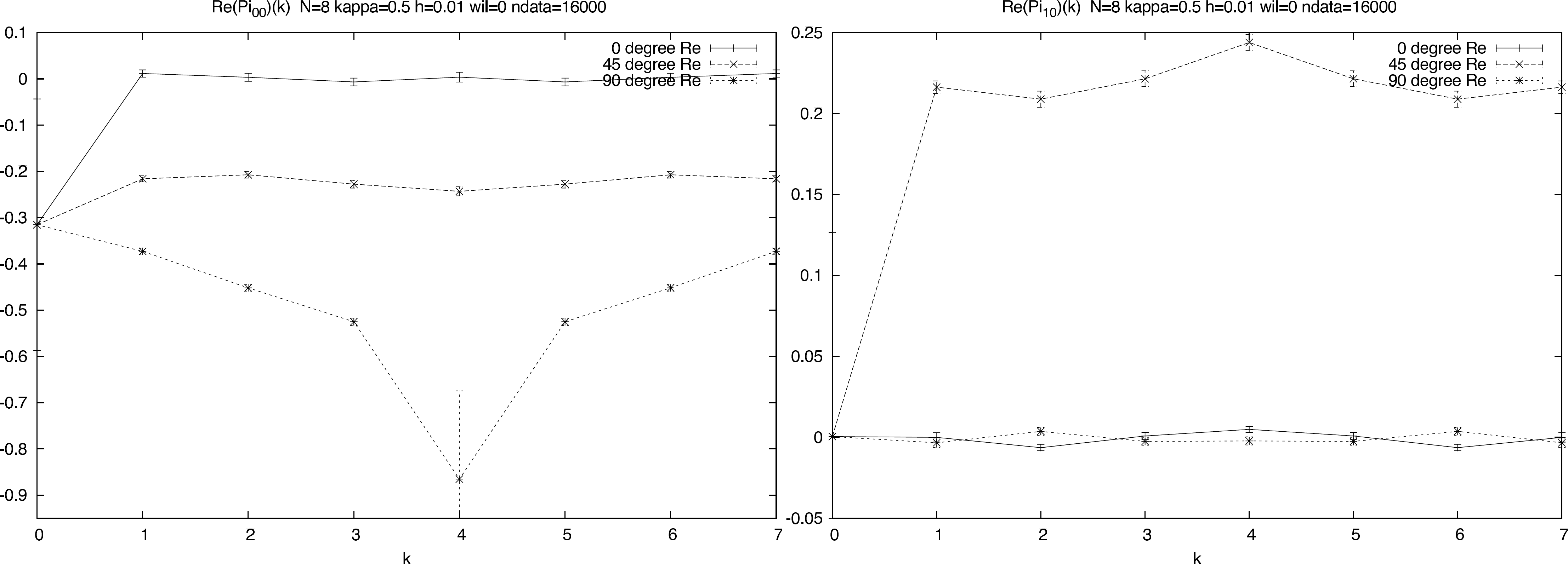}
\caption
{The real parts of $\Pi_{00}$ and $\Pi_{10}$ of the mirror for $\kappa = 0.5$, $h=0.01$: symmetric phase, three massless chiral fermions, see Section 5; the dip at $k=4$ is due to an ``Umklapp" process (an analytic calculation using the mirror partition function $\rm{det}_+(2 - D)$ gives a very similar result).
}
 \label{p_0.5_0.01} 
\end{figure}
At $h=0$,  in Section \ref{anotherview}, we give  analytical arguments that the massless spectrum should contain three charged massless chiral fermions (two of them forming a vectorlike pair and thus not contributing to the anomaly). This is borne out by the simulations done in the $h\rightarrow 0$ limit. For example, the value of $|2 C|$ inferred from the plot on Fig.~\ref{p_0.5_0.01} of $\tilde\Pi_{11}(90^0)$ is roughly $.39$, while that inferred from $\tilde\Pi_{12}(45^0)$ it is $.44$ (closer to the continuum value of $3/(2 \pi) \simeq .48$). A calculation (which for brevity we do not show) of the polarization operator of   the mirror partition function $\rm{det}_+(2 - D)$, see eqn.~(\ref{Ztotal2}) in  Section \ref{anotherview}, also shows good agreement with the $h\rightarrow 0$ Monte Carlo data shown on Fig.~\ref{p_0.5_0.01}.

\subsubsection{Scaling of the different contributions to $\Pi_{\mu \nu}$ as $\kappa \rightarrow 0$}
\label{MCresultskappa}

\begin{figure}
\begin{center}
\includegraphics[width=5in]{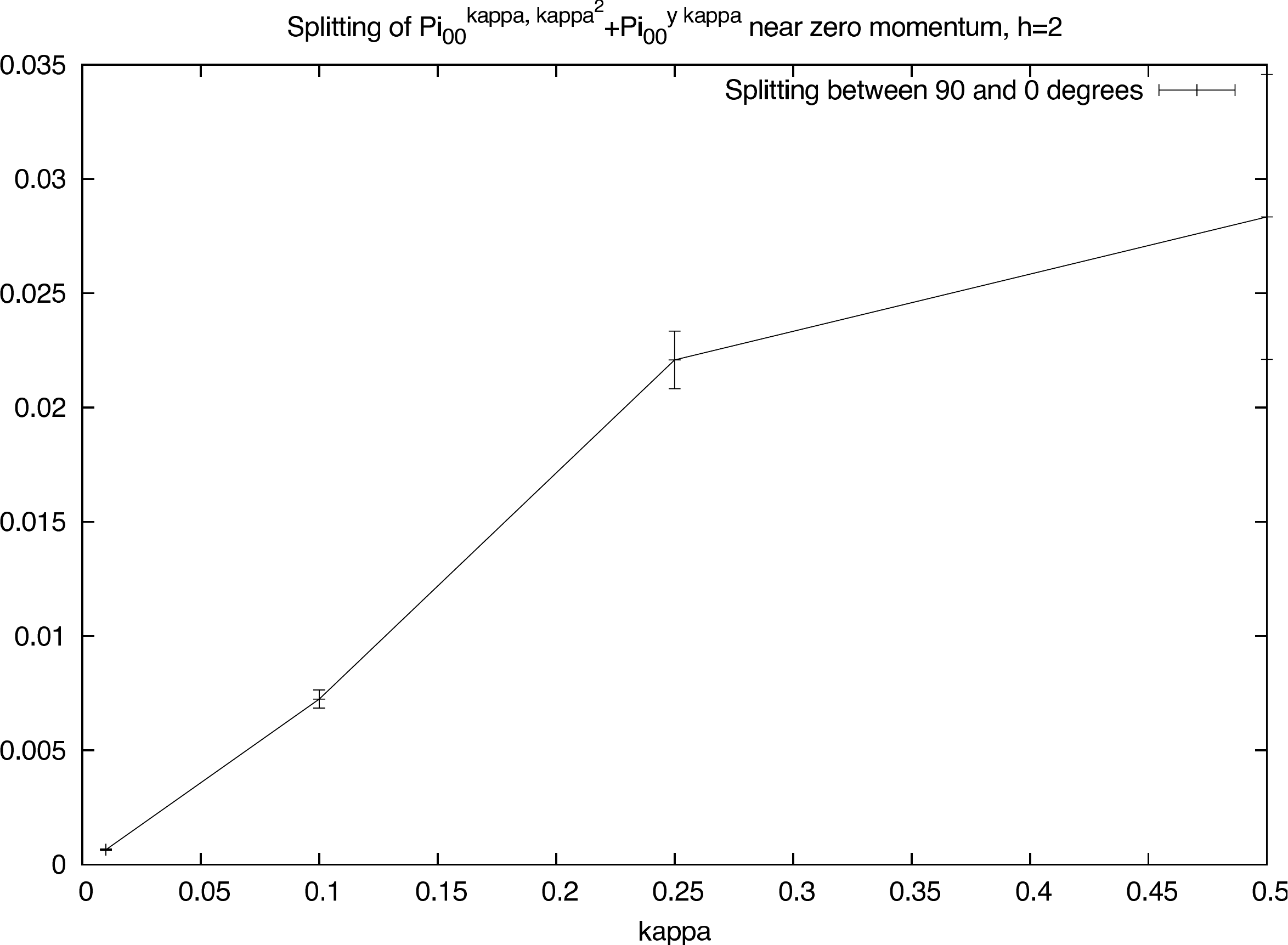}
\caption
{The small-$\kappa$ scaling of   $(\Pi_{11}^\kappa + \Pi_{11}^{y\kappa})\big\vert^{90^0}_{0^0}$ of the $\kappa$-dependent terms' contribution to the discontinuity of the mirror polarization operator at small momentum.
}
 \label{kappalim} 
\end{center}
\end{figure}

Finally, on Fig.~\ref{kappalim} we show the $\kappa$-scaling of the $\kappa$-dependent terms in $\Pi_{\mu\nu}$ of the mirror theory (i.e. deep into the symmetric phase).
More precisely, we show the $0$-$90$ degree split of the sum of the $\Pi_{00}^{\kappa}$ (\ref{zmirror13}) and $\Pi_{00}^{y\kappa}$  (\ref{zmirror14}) at small momenta as a function of $\kappa$. It is clear that the contribution of the $\kappa$-dependent terms scales to zero with $\kappa$, showing that in the symmetric phase contributions to the angular discontinuity of polarization operator come from the Yukawa terms, $\Pi^{y, y^2}$ of eqn.~(\ref{zmirror9}).
\bigskip

\section{Attempt at  (some) analytic understanding}
\label{anotherview}

In this Section, we will attempt to get a somewhat better analytic understanding of the numerical results of the previous Sections. 
To this end, we will develop a different representation of the 1-0 model partition function.
We begin by noting that the 1-0 model action (\ref{toymodel}) can be written in unprojected components (the superscripts $\hat{P}^{1 (0)}$ in the hatted projectors indicate whether the charge-1 or charge-0 Dirac operator has to be used):
\begin{eqnarray}
\label{toymodel2}
S_{kinetic}&=& - \left( \bar\psi   D_1     \psi \right) - \left( \bar\chi    D_0    \chi \right) \nonumber  \\
 S_{Yukawa} &=&  y \left\{ \left( \bar\psi \hat{P}_+^1   \phi^*  P_+  \chi  \right) + \left( \bar\chi \hat{P}_-^0    \phi {P}_- \psi  \right)   + h \left[ \left( \psi^T (P_-)^T  \phi \gamma_2   P_+  \chi \right) - \left( \bar\chi \hat{P}_-^0  \gamma_2  \phi^* (\hat{P}_+^1)^T  \bar\psi^T \right) \right] \right\} \nonumber~,
\end{eqnarray}
where the measure now is the usual vectorlike theory measure in terms of $\psi, \bar\psi, \chi, \bar\chi$:
\begin{eqnarray}
\label{Ztotal}
Z = \int d \psi d \bar\psi d \chi d\bar\chi d \phi \; e^{- S_{kinetic} - S_{Yukawa} - S_\kappa} ~,
\end{eqnarray}
and $S_\kappa$ is defined in (\ref{Skappa}). 
We now perform the field redefinition:
\begin{eqnarray}
\label{fieldredef1}
\psi  \rightarrow \psi, && \bar\psi \rightarrow \bar\psi\; {1 \over 2 - D_1}, \nonumber \\
\chi \rightarrow \chi, && \bar\chi \rightarrow \bar\chi \;{1 \over 2 - D_0}.
\end{eqnarray}
The motivation for this redefinition can be traced back to the GW relation, which  implies that:
\begin{equation}
\label{GW10}
{1 \over 2 - D}\; \hat{\gamma}_5 = \gamma_5 \;{1 \over 2 - D} ~,
\end{equation}
in other words, a $\hat{\gamma}_5$  action on the original $\bar\psi$ is transformed into the action of  $\gamma_5$ on the new fields in (\ref{fieldredef1}).
The price to pay for having an action of the lattice chiral symmetries generated by $\gamma_5$, exactly as in the continuum, is the nonlocality of the redefinition (\ref{fieldredef1}) and thus of the resulting action (see (\ref{Ztotal1}) below).

Since we  work  perturbatively  in the gauge field and at finite volume, we will  imagine throughout this Section that the singularity at $D=2$ of (\ref{fieldredef1}) is  avoided by turning on background Wilson lines. The effect of the field redefinition on the partition function is:
\begin{eqnarray}
\label{Ztotal1}
Z &=& \int d \psi d \bar\psi d \chi d\bar\chi d \phi \; {\rm det}(2 - D_1) \; {\rm det}(2 - D_0) \; e^{ -S_{kinetic}^\prime - S_{Yukawa}^\prime - S_\kappa}~,\nonumber \\
S^\prime_{kinetic} &=&  - \left( \bar\psi  \; {D_1 \over 2 - D_1}  \;  \psi \right) - \left( \bar\chi  \;  {D_0 \over 2 - D_0}   \; \chi \right)~,\\ 
S^\prime_{Yukawa} &=&   {y\over 2}\left( \bar\psi    \phi^*  P_+  \chi  \right) + {y\over 2}\left( \bar\chi      \phi {P}_- \psi  \right)   \nonumber \\
&+& yh \left[ \left( \psi^T    \phi \gamma_2   P_+  \chi \right) - {1 \over 4} \left( \bar\chi   P_-  \gamma_2  \phi^* \bar\psi^T \right)    - {1 \over 4} \left( \bar\chi P_- {D_0 \over 2 -D_0}  \gamma_2  \phi^* ( {D_1 \over 2 - D_1})^T P_+   \bar\psi^T \right) \right]    \nonumber~.
\end{eqnarray}
Obtaining the transformed Yukawa couplings in $S_{Yukawa}^\prime$ requires repeated use of the GW relation  and the equivalent relation (\ref{GW10}). Needless to say, similar redefinitions hold in four dimensions and can be straightforwardly performed if necessary. 

Several comments, concerning the representation (\ref{Ztotal1}) of the mirror partition function, are now in order. We hope that these comments are useful to clarify the relation between different formulations of exact lattice chirality:
\begin{itemize}
\item
 The singularity of the action at  the position of the doublers, $D=2$, ensures that they have  infinite action and do not propagate at the classical level,  as first proposed by Rebbi \cite{Rebbi:1986ra}. The problem with \cite{Rebbi:1986ra}, pointed out in  \cite{Bodwin:1987gi, Pelissetto:1987ad}, of the doublers contributing as ghosts to  the photon polarization operator at the quantum level is solved by the determinant prefactors, which exactly cancel the would-be ghost/doubler contributions.
 \item Another comment concerns the relation of (\ref{Ztotal1}), with $y = h =0$, to  domain wall fermions. In the 2d case, these propagate on a finite interval in three dimensions. The generating functional of Green's functions of the boundary chiral modes---in the case at hand,   one charged and one neutral---can be obtained by integrating out  the bulk fermions. This is  technically possible since the fermion action is bilinear and there is no gauge field propagation in the extra dimension (or any other non-uniformity except at the boundaries). When taking the chirally-symmetric limit of an infinite number of sites in the extra dimension,  a massive bulk Pauli-Villars field, antiperiodic in the extra dimension, has to be included in order to obtain a finite determinant. The generating functional of Green's function for the boundary chiral modes can be represented as a partition function with source terms. The result is exactly (\ref{Ztotal1}), with $y = h =0$, and with source terms for $\psi$ and $\chi$ included. The determinant prefactors in (\ref{Ztotal1}) arise as a combination of the determinants of the bulk fermions and the Pauli-Villars fields. The derivation of these results   can be extracted  from the work of ref.~\cite{Kikukawa:1999sy} (a less rigorous but nonetheless enlightening continuum discussion of the properties of the kinetic operator for the boundary modes is given in \cite{Fosco:2007ry}).
\item For vanishing Majorana coupling, $h=0$, the   Yukawa interaction in (\ref{Ztotal1}) is equivalent to the chirally invariant Yukawa coupling of ref.~\cite{Luscher:1998pq}, see also \cite{Chiu:1998aa}. To see this, 
use $(\bar\psi {D \over 2 - D} \psi) = {1 \over 2} (\bar\psi D \psi) + {1 \over 2} (\bar\psi D {D \over 2 - D} \psi)$  to replace the $\psi$ kinetic term in the action $S_{kin}^\prime$. Then note that $- {1 \over 2} (\bar\psi D {D \over 2 - D} \psi) = (\bar\xi (2 - D) \xi) + {1 \over \sqrt{2}} (\bar\xi D \psi) + {1 \over \sqrt{2}} (\bar\psi D \xi)$, where it is understood that the new charged field $\xi$ is integrated out from the action using its equation of motion, while the $\xi$-determinant exactly cancels  the one in (\ref{Ztotal1}). Then, shift the integration variable $\psi \rightarrow \psi + \sqrt{2} \xi$. Next, perform exactly the same operations on $\chi$, introducing a neutral field $\eta$, to finally obtain the  action in the original  form of L\" uscher, suitably adapted to the 2d case and to our normalization:
\begin{equation}
\label{luscheryukawa} S = - {1\over 2}(\bar\psi D_1 \psi + \bar\chi D_1 \chi) + 2 \bar\xi \xi + 2 \bar\eta \eta  + {y \over 2}  (\bar\psi + \sqrt{2} \xi) \phi^* P_+ (\chi + \sqrt{2} \eta) +   {y \over 2}(\bar\chi + \sqrt{2} \eta) \phi P_- (\psi + \sqrt{2} \xi),
\end{equation}
where the measure is the trivial one over $\psi, \chi, \eta, \xi$.  The Yukawa interaction of (\ref{toymodel}) is thus equivalent to that of \cite{Luscher:1998pq}. We note that gauge and chiral invariant Majorana couplings were not considered   in \cite{Luscher:1998pq} and to the best of our knowledge were first constructed in \cite{Giedt:2007qg}.
\end{itemize}

  We can now use the representation (\ref{Ztotal1}) to split the partition function exactly as we did in the basis of GW fermions. The representation of the split partition function that we give below is, in fact, equivalent to that in (\ref{z01}).
  The splitting of the $\psi$-$\chi$ partition function (\ref{Ztotal1}) into a ``light" and ``mirror" part is now done trivially using the $\gamma_5$-eigenvectors, which have no gauge-field dependence. We obtain, denoting now by $\psi_\pm, \chi_\pm$ the ``normal" $\gamma_5$-chirality components of the 2-component $\psi,\chi$, and using the fact that $S^\prime_{Yukawa}$ only depends on the mirror components of $\psi, \chi$:
  \begin{eqnarray} \label{Ztotal2}
  Z &=& Z_+ \times Z_- \times {1 \over J} ~,\nonumber \\
  Z_+&=& {\rm det}_- (2 - D_0)  {\rm det}_+ (2 - D_1) \int d \psi_+ d \bar\psi_+ d \chi_- d\bar\chi_- e^{- \left( \bar\psi_+    {D_1 \over 2 - D_1}    \psi_+ \right) - \left( \bar\chi_-   {D_0 \over 2 - D_0}    \chi_- \right)  }  ~,\\ 
  Z_-&=& {\rm det}_+ (2 - D_0) {\rm det}_- (2 - D_1) \int d \psi_- d \bar\psi_- d \chi_+ d\bar\chi_+ d \phi  e^{- \left( \bar\psi_-    {D_1 \over 2 - D_1}   \psi_- \right) - \left( \bar\chi_+   {D_0 \over 2 - D_0}    \chi_+ \right)  - S^\prime_{Yukawa} - S_\kappa}  ~.\nonumber
\end{eqnarray}
Splitting the determinant prefactor into ``light" and ``mirror," as indicated in (\ref{Ztotal2}),   requires  using the gauge-field dependent eigenvectors of $\hat{\gamma}_5$:
\begin{eqnarray} \label{Ztotal3}
{\rm det}_+ (2 - D) &=&{\rm det}_+ (1 + \hat{\gamma}_5 \gamma_5)  =  {\rm det} || (u^\dagger_i (2-D)t_j)|| =  {\rm det} ||2 (u^\dagger_i t_j)|| , \\
{\rm det}_- (2 - D) &=& {\rm det}_- (1 + \hat{\gamma}_5 \gamma_5)=   {\rm det} ||  (w^\dagger_i (2 - D)v_j)||= {\rm det} ||2 (w^\dagger_i v_j)|| , \nonumber
\end{eqnarray}
 where the appropriate $\hat{\gamma}_5$-eigenvectors are to be used for $D^0$ or $D^1$.  The gauge variation of the mirror partition function $Z_-$ has now two contributions: one from the variation of ${\rm det}_- (2 - D^1)$  and one from the variation of the path integral over $\psi_-, \chi_+, \phi$. The splitting theorem, applied to the chiral partition function defined by the latter, can be easily seen to imply that the gauge variation   $Z_-$ with  the determinant factors left out vanishes. Thus the entire gauge variation of $Z_-$ comes from the determinants:
\begin{eqnarray}
\label{vartn1}
\delta \ln Z_- &=& \delta_\omega {\rm ln det} ||  (w^\dagger_i (2-D^1)v_j)|| =  \sum_j (\delta_\omega w_j^\dagger  w_j ) + {\rm tr}\; \hat{P}_+ {1\over 2 - D^1} \delta_\omega(2- D^1) \nonumber \\
&=&   \sum_j (\delta_\omega  w_j^\dagger w_j ) + i  {\rm Tr}\omega \left(  \hat{P}_+^1 - P_+ \right) =  \sum_j ( \delta_\omega w_j^\dagger w_j ) + {i \over 2} {\rm Tr} \omega \hat{\gamma}_5 , 
\end{eqnarray}
where, as usual, the measure current is cancelled by the variation of the Jacobian and light partition function. The gauge variation of $Z_-$ is, naturally, the same as in (\ref{z03}) (the gauge variation of $Z_+$ can be obtained similarly to (\ref{vartn1}) and be seen to combine, together with the Jacobian to cancel that of $Z_-$).
 
 The determinant prefactor in $Z_-$ contributes both to the real and imaginary parts of the mirror polarization operator. It is clear from (\ref{vartn1}) above that the imaginary part of the polarization operator due to the determinant is exactly as required by anomaly cancellation.  Assuming   unitarity at long distances, one would  argue that Re$\Pi_{\mu\nu}$ should  receive a contribution from at least one massless charged chiral fermion (plus, possibly, a number of massless states in anomaly-free representations).   The goal of our simulation was to precisely find out the real part of the mirror polarization operator. Let us now compare the numerical findings with what can be inferred from the representation of the mirror partition function (\ref{Ztotal2}, \ref{Ztotal3}).  
 \begin{enumerate}
 \item
The real part of the mirror theory $\Pi_{\mu\nu}^-$ receives two contributions at $y=\infty$. The first is due to   the determinant prefactor. The contribution to Re$\Pi_{\mu\nu}^-$ of ${\rm det}_- (2 - D^1)$ is easily seen to be exactly that of   three massless propagating charged chiral fermions---the chiral components of the three 2d doubler modes. This can be easily done using the formulae from the Appendix to calculate the determinant contribution to $\Pi_{\mu\nu}$, then numerically plotting the result and comparing to the long-distance contribution of a single charged chiral fermion. The determinant prefactor contribution to Re$\Pi_{\mu\nu}^-$ is, obviously, the same for any value of $y, h$.
\item The only other  contribution to Re$\Pi_{\mu\nu}^-$ at $y=\infty$ arises 
from the nonlocal coupling in the Majorana Yukawa term in $S_{Yukawa}^\prime$  of (\ref{Ztotal1}). While it appears difficult to calculate analytically this contribution in the disordered-$\phi$ phase, 
our numerical results for the mirror polarization operator Re$\Pi_{\mu\nu}$  show  that there is {\it one} massless propagating chiral fermion. 
Thus  two of the three massless modes contributed by the prefactor are cancelled by the nonlocal contribution to Re$\Pi_{\mu\nu}$ from the Majorana term in $S_{Yukawa}$; it appears that this cancellation is exact for all values of $h>1$.
\item
For vanishing Majorana coupling, $h=0$, on the other hand, both   numerical simulations (with the code of \cite{Giedt:2007qg} used in this paper, we  can only approach the $h \rightarrow 0$ limit) and analytic arguments using the representation (\ref{Ztotal2}), to be discussed in more detail elsewhere,  show that the real part of the mirror polarization operator at small momenta and infinite $y$ is that of three charged massless chiral fermions. This is not too surprising as the Majorana coupling goes away in the limit and at infinite $y$ the only dependence on the gauge field is in the determinant  prefactor in $Z_-$ of (\ref{Ztotal3}).
This indicates the  crucial role of the Majorana-type couplings (recall that they were motivated by the need to   break all mirror global symmetries) in facilitating the decoupling of the maximal possible number (allowed by anomaly matching) of charged mirror degrees of freedom.
\item Another lesson we learned  is that probing the fermion spectrum with local fermion operators (including charged local fermion-scalar composites, as in \cite{Giedt:2007qg}) can miss massless degrees of freedom. The massless charged mirror fermions   were not seen in that study, perhaps because they are not  expressed in an obvious  local way through the original variables. The long-distance gauge-boson polarization operator of the mirror theory is a universal probe of the charged mirror spectrum and should  be the first quantity, along with susceptibilities probing  chiral symmetry breaking,  computed in any future studies of anomaly-free models.
\end{enumerate} 
      
   {\flushleft
{\bf \large Acknowledgements}}

\vspace{5pt}
We thank Maarten Golterman and Yigal Shamir for valuable correspondence and insightful comments which led to this work, and for comments on the manuscript. We also thank Joel Giedt for comments on the manuscript and Siavash Golkar for discussions. We acknowledge support by the National Science and Engineering 
Research Council of Canada. We thank the Canadian Institute for Theoretical Astrophysics (CITA) for giving us the opportunity to use their computer cluster.

\appendix

\section{More fun with chiral polarization operators}
\label{polarizationops}

Consider first 
the mirror polarization operator (\ref{pimirror2}) at $y=0$, which can be written as $\Pi^-_{\mu\nu}= \delta_\nu j_\mu^w + \Pi^{\prime -}_{\mu \nu}$.
The $\Pi^{\prime \pm}_{\mu\nu}$ part of the  polarization operator is independent of the basis vectors and can be expressed (easily verified by looking at (\ref{pimirror2})) as:
\beq
\label{piprime1}
\Pi^{\prime \pm}_{\mu\nu} = \delta_\nu \Pi^{\prime \pm}_\mu~,
\eeq
with:
\beq
\label{piprime2}
\Pi^{\prime \pm}_\mu = {\rm Tr} P_\pm D^{-1} \delta_\mu D = {1\over 2} {\rm Tr} D^{-1} \delta_\mu D \pm {1 \over 2} {\rm Tr} \gamma_5 D^{-1} \delta_\mu D~.
\eeq
We will denote the two terms in (\ref{piprime2}) as the vector (superscript $V$) and axial (superscript $5$) parts of $\Pi_\mu^\prime$:
\beqa
\label{piprime3}
\Pi_\mu^V &\equiv& {\rm Tr} D^{-1} \partial_\mu D \nonumber \\
\Pi_\mu^5&\equiv& {\rm Tr} \gamma_5 D^{-1} \partial_\mu D~.
\eeqa
We note that $\Pi_\mu^V$ is real, since $D^\dagger = \gamma_5 D \gamma_5$.
On the other hand, we easily obtain (using the GW relation on the way) that:
\beqa
\label{piprime4}
{\rm Re}\; \Pi_\mu^5 &=& {1 \over 2} {\rm Tr} \gamma_5 \delta_\mu D \nonumber \\
i \;{\rm Im} \; \Pi_\mu^5 &=& {1 \over 2 } {\rm Tr} \left[ \delta_\mu D, \gamma_5 \right] D^{-1} ~.
\eeqa
Thus, we can express $\Pi_{\mu\nu}^{\prime \pm}$ (\ref{piprime1}) as follows:
\beq
\label{piprime5}
\Pi^{\prime \pm}_{\mu \nu} = {1 \over 2}\; \delta_\nu {\rm Tr} D^{-1} \delta_\mu D \pm {1\over 4} \;{\rm Tr} \gamma_5 \delta_\mu \delta_\nu D \pm {1\over 4}\; \delta_\nu {\rm Tr} \left[ \delta_\mu D, \gamma_5 \right] D^{-1}
\eeq
A few comments:
\begin{enumerate}
\item 
The first two terms of $\Pi_{\mu\nu}^{\prime \pm}$ are real and symmetric in $\mu,\nu$, while the last term is purely imaginary and has both a symmetric and an antisymmetric part. 
\item
The first term,  ${1 \over 2} \delta_\nu {\rm Tr} D^{-1} \delta_\mu D$, in (\ref{piprime5}) is simply $1/2$ the vector theory polarization operator. It is manifestly real and has a nonlocality (a factor of $D^{-1}$) due to the fact that a massless particle is propagating in the loop. It is also manifestly transverse w.r.t. both indices. In the continuum limit it  reduces to the transverse Lorentz invariant contribution  (the first term in (\ref{fermion})) with a coefficient equal to one-half that of the massless Schwinger model.
\item The second term, $ \pm {1\over 4} {\rm Tr} \gamma_5 \delta_\mu \delta_\nu D$, is manifestly symmetric and is  local (with the usual exponential tail). It is transverse w.r.t.~both $\mu$ and $\nu$.
This term should vanish in the continuum limit, as there is no symmetric, local, transverse, parity-odd expression one can write in the continuum. In fact, numerical evaluation of the trace on finite lattices indicates that this term vanishes identically at $A=0$; we have no analytic proof, but feel that  one should exist. 
\item The last term, $\pm {1\over 4} \delta_\nu {\rm Tr} \left[ \delta_\mu D, \gamma_5 \right] D^{-1}$, is purely imaginary, and has no manifest symmetry in $\mu, \nu$. We know from general arguments  that its antisymmetric part should equal ${1 \over 2} {\cal F}_{\mu\nu}$ (see also (\ref{eq:mysterious}) below) and thus be local. Its symmetric part, on the other hand, is nonlocal. The divergences of the symmetric and antisymmetric parts are equal,  each giving rise to one-half the anomalous divergence of $\Pi_{\mu\nu}^{\prime \pm}$. Once again, this is easy to also explicitly check from (\ref{piprime5}):
\beq
\label{piprime6}
\nabla^*_\mu \Pi^{\prime \pm}_{\mu \nu}(x,y) = \pm {1 \over 4} \nabla^*_\mu \; \delta_\nu {\rm Tr} \left[ \delta_\mu D, \gamma_5 \right] D^{-1} = \mp {i \over 2}{\delta \over \delta A_\nu(y)}\; {\rm tr}\;\hat{\gamma}_5 (x,x)~.
\eeq 
\end{enumerate}

{\flushleft T}wo questions are to be answered in order to generalize these results to the most general chiral 
theories (e.g.~to our $y\ne0$ mirror).  
Our object of interest $\Pi^{- \; \prime}_{\mu\nu}$ is again defined by:
\begin{equation}
\label{eq:def}
\Pi^-_{\mu\nu}=\delta_\nu j^w_\mu+\Pi^{- \; \prime}_{\mu\nu}~.
\end{equation}
First, is it true that $\Pi^{- \; \prime}_{\mu\nu}$ is always a total derivative?  Namely, does there always exist a $\Pi^{- \; \prime}_\mu$ such that:
\[\Pi^{- \; \prime}_{\mu\nu}=\delta_\nu \Pi^{- \; \prime}_\mu\,?\]
Second, if such a $\Pi^{- \; \prime}_\mu$ does exist, is it manifestly gauge invariant?

The answer to the first question is immediately yes if one just looks at the definition 
of $\Pi^{- \; \prime}_{\mu\nu}$ given by equation \eqref{eq:def} and recalls that both
$\Pi^-_{\mu\nu}\equiv\delta_\mu\delta_\nu\log Z_-$ and $\delta_\nu j^w_\mu$ are total
derivatives.  In fact, we know precisely what $\Pi^{- \; \prime}_\mu$ is.
Given any chiral action, in the notation of \cite{Poppitz:2007tu}:
\[\tilde S[X, Y^\dag, O]\equiv \exp(S[X, Y^\dag, O])\, , \]  
and the partition function:
\begin{equation}
Z\equiv \int \Pi \ud \bar c \ud c \, \tilde S[c_i u_i, \bar c_i v^\dag_i, O]~,
\end{equation}
with $u_i$ and $v_i$ some appropriate eigenvectors, 
we have proved the ``splitting-theorem'':
\begin{equation}
\label{eq:first_derivative1}
\delta_\mu \log Z= j_\mu+\frac{Z^{(1)}_\mu}{Z}~,
\end{equation}
where:
\begin{equation}
Z^{(1)}_\mu\equiv\int \Pi\ud \bar c\ud c\, \delta_{O,\mu}\tilde S[c_i u_i, \bar c_i v^\dag_i, O].
\end{equation}
Here $\delta_{O,\mu}\tilde S$ represents the variation of $\tilde S$ with respect to $A_\mu$
with the vectors $u_i$ and $v_i$ fixed (namely, the variation of $\tilde S$ due to 
the variations of the operators only).  As explained in Section \ref{setup1}, to make sure
$\delta_{O,\mu}\tilde S[c_i u_i, \bar c_i v^\dag_i, O]$ is again chiral, one should
insert additional $P$'s if needed, and the value of $Z^{(1)}_\mu$ remains intact in this
procedure.  We will implicitly assume this rule here.
Comparing equation \eqref{eq:first_derivative1} with equation \eqref{eq:def} implies:
\begin{equation}
\Pi^{- \; \prime}_{\mu\nu}=\delta_\nu \left(\frac{Z^{(1)}_{- \; ,\mu}}{Z_-}\right).
\end{equation}

Now, the answer to the second question is also yes provided that $\tilde S$ is gauge
invariant.  We assume that the action $\tilde S[X, Y^\dag, O]$ is invariant under 
the gauge transformation
\begin{equation}
\delta X=i\omega X, \qquad \delta Y=i\omega Y, \quad \textrm{and}\quad
\delta O=i\commut{\omega}{O}.
\end{equation}
It is easily verified that $\delta_{O, \mu}\tilde S$ is also gauge invariant, where
all the operators, including $\delta_\mu O$ and the additional $P$'s inserted, 
transform as adjoints as well.   Therefore both $Z$ and $Z^{(1)}_\mu$ are 
partition functions defined by a ``chiral'' and gauge invariant
action.  As given by equation (4.55) in \cite{Poppitz:2007tu}, their gauge 
transformations are completely determined independent to the actual 
expressions of the actions.  Indeed, 
\begin{equation}
\delta_\omega \log Z=\delta_\omega\log Z^{(1)}_\mu= j_\omega-i\Tr\omega(\hat P-P)~,
\end{equation}
where $j_\omega$ is the measure current corresponding to an infinitesimal gauge transformation with parameter $\omega$. 
Thus, we find:
\begin{equation}
\delta_\omega \log\frac{Z^{(1)}}{Z}=\delta_\omega\log Z^{(1)}-\delta_\omega\log Z=0,
\end{equation}
which says that $\log\frac{Z^{(1)}}{Z}$ is gauge invariant, and so is $Z^{(1)}\over Z$.

Finally, we derive the general properties of chiral polarization operators listed in Section \ref{generalpiproperties}. 
Gauge invariance of $\Pi^{- \; \prime}_\mu$ implies immediately, that in a general classically gauge invariant chiral theory:
\begin{equation}
\label{divsecond}
\nabla^\ast_\nu\Pi^{- \; \prime}_{\mu\nu}=0, 
\end{equation}
following the steps that led from (\ref{z9}) to (\ref{z11}). 
Furthermore, in Section \ref{transversality} we found that  the divergence of $\Pi^{- \; \prime}_{\mu\nu}$  with respect to the first index $\mu$
is  independent of the action:
\begin{equation}
\label{abc}
\nabla^\ast_\mu \Pi^{- \; \prime}_{\mu\nu}=\frac{i}{2}\delta_\nu \tr \hat \gamma^5_{xx}.
\end{equation}
Another action-independent result is that the anti-symmetric part of $\Pi^{- \; \prime}_{\mu\nu}$ is 
also known explicitly---since $\Pi^-_{\mu\nu}$ is manifestly symmetric, the antisymmetric parts of $\Pi^{- \; \prime}_{\mu\nu}$ and $\delta_\nu j_\mu^w$  should cancel:
\begin{equation}
\label{eq:mysterious}
\Pi^{- \; \prime \; A}_{\mu\nu}\equiv \frac{1}{2}(\Pi^{- \; \prime}_{\mu\nu}-\Pi^{- \; \prime}_{\nu\mu})
=\frac{1}{2}\mathcal{F}_{\mu\nu},
\end{equation}
where,
\begin{equation}
\mathcal{F}_{\mu\nu}=\delta_\mu  j^w_\nu-\delta_\nu j^w_\mu
=-\Tr(\hat P_- \commut{\delta_\mu\hat P_- }{\delta_\nu\hat P_- })
\end{equation}
is manifestly local.
Its divergence, after a few steps of algebra, turns out to be 
$\frac{i}{2}\delta_\nu \tr\hat\gamma^5_{xx}$ as well. 
 Applying (\ref{identity}) to $\mathcal F_{\mu\nu}$
and using the gauge transformation $\delta_\omega\hat P_-=i\omega\hat P_--i\hat P_-\omega$,
we find:\footnote{We used identities $\hat P_-\delta \hat P_-\hat P_-=0$ and 
$\delta\hat P_-\hat P_-+\hat P_-\delta \hat P_-=\delta\hat P_-$, both from
$\hat P_-^2=\hat P_-$.}
\begin{equation}
\begin{split}
\nabla^\ast_\mu \mathcal F_{\mu\nu}=&
\; i \,\frac{\delta}{\delta\omega}
\Tr[\delta_\nu\hat P_- \hat P_-(\omega \hat P_--\hat P_-\omega)
		-\hat P_-\delta_\nu\hat P_-(\omega \hat P_--\hat P_-\omega) ]\\
=&\; i\,\frac{\delta}{\delta\omega}\Tr\delta_\nu\hat P_-\omega
=\;\frac{i}{2}\delta_\nu \tr\hat \gamma^5_{xx}~.
\end{split}
\end{equation}
Combining these results  leads to:
\begin{equation}
\label{eq:half_div}
\nabla^\ast_\mu\Pi^{- \; \prime \; A}_{\mu\nu}=\frac{1}{2}\nabla^\ast_\mu \Pi^{- \; \prime}_{\mu\nu}
\end{equation}
In other words, the anti-symmetric  and symmetric parts of $\Pi^{- \; \prime}_{\mu\nu}$
  each contribute  half of the anomalous divergence. 
  The results (\ref{divsecond}), (\ref{abc}), (\ref{eq:half_div}) just derived hold for general chiral partition functions, independent of the action, and in particular for our $y\ne 0$ mirror theory. Verifying that these exact properties  hold 
 is an important check on any numerical simulation.
 
\section{The Neuberger-Dirac operator and its perturbative expansion}
\label{notation}

The Wilson operator in our convention is given by: 
\beqa
\label{wilson1}
X_{mn} &=& M \delta_{\mu,\nu} + {1 \over 2}\; \sum_\mu   - \gamma_\mu \left( \delta_{m + \hat{\mu}, n} U_\mu(m) - \delta_{m, n+\hat{\mu}} U^\dagger_\mu(n) \right)   \nonumber \\ 
&+&   {1 \over 2}\; \sum_\mu \; r\; \left( 2 \delta_{mn} - \delta_{m+\hat{\mu},n} U_\mu(m) - \delta_{m, n+\hat{\mu}} U^\dagger_\mu(n)  \right)~,
\eeqa
where $m,n$ label $d$-dimensional lattice sites, $U_\mu(m) = e^{i A_\mu(m)}$, $\hat{\mu}$ is a unit vector in the $\mu$-th direction on the lattice, and $\delta_{m,n}$ is a $d$-dimensional Kronecker symbol. The Wilson operator (\ref{wilson1}) is $\gamma_5$ hermitean $(X^\dagger)_{mn} = \gamma_5 X_{mn} \gamma_5$. We define $\hat{\gamma}_5$ as follows:
\beq
\label{gamma51}
\hat{\gamma}_5= {1 \over \sqrt{ X X^\dagger}} X, ~~ \hat{\gamma}_5^2 = 1~, 
\eeq
in terms of which our convention for the GW operator is:
\beq
\label{gwoperator}
D = 1 - \hat{\gamma}_5 \gamma_5  = D_0 + D_1 + D_2 + \ldots~, 
\eeq
where $D_{0,1,2}$ are the terms in the expansion of $D$ around the trivial gauge background. We take $M=1$, $r = -1$ from now on. 
We define our finite-volume Fourier transforms as follows, using $\omega_N= e^{2 \pi i \over N}$:
\beq
\label{FT1}
X(q,p) = {1\over N^{d\over 2}} \sum_{m,n} \omega_N^{- q\cdot m + p \cdot n} X_{mn}~. 
\eeq
We also define the functions:
\beq
\label{sc}
s(p) = \sin { \pi p \over N}~, ~~c(p) = \cos { \pi p \over N}~.
\eeq
By $\sigma_0$, we denote the $2^{d \over 2}$-dimensional unit matrix; in d=2, we use $\gamma_1 = \sigma^1, \gamma_2 = \sigma^2, \gamma_5 = \sigma^3$, where $\sigma^i$ are the Pauli matrices.
Similar to (\ref{gwoperator}), we also have an  expansion of $X$ in powers of $A_\mu$:
\beq
\label{xexpansion}
X(p,q) = X_0 (p,q) + X_1(p,q) + X_2 (p,q) + \ldots, 
\eeq
where the trivial-background term is given by:
\beq
\label{x0}
X_0(p,q) =  \delta_{p,q} x_0(p), ~ {\rm where }~ ~x_0(p) \equiv \left( 1 - \sum_\mu 2 s^2(p_\mu)\right) \sigma_0 - 
\sum_\mu i \gamma_\mu s(2 p_\mu) ~.
\eeq
The linear and quadratic terms of (\ref{xexpansion}) are:
\beqa
\label{x12}
X_1(p,q) &=&- {1\over N^2} \sum_\mu \tilde{A}_\mu(p-q)\; \omega_N^{q_\mu - p_\mu \over 2} \left[ i \gamma_\mu c(p_\mu+q_\mu) +\sigma_0 s(p_\mu + q_\mu) \right] \;, \\
X_2(p,q) &=& -{1\over 4 N^2} \sum_\mu\left[ (1 - \gamma_\mu)\; \omega_N^{q_\mu} \sum_m (A_\mu(m))^2 \;\omega_N^{(q - p) \cdot m} + (1 + \gamma_\mu) \; \omega_N^{- p_\mu} \sum_{n} (A_\mu(n))^2 \;\omega_N^{(q - p) \cdot n}  \right] \nonumber
\eeqa
where in $X_1$ we use  the Fourier transform of $A_\mu$, defined by:
\beq
 \label{ftA}
 \tilde{A}_\mu(p) = \sum_m \omega_N^{- p\cdot m} A_\mu(m).
 \eeq
In terms of $X_0, X_1, X_2$ the various terms in the expansion of the GW operator (\ref{gwoperator}) are described below. We begin with 
the free GW operator, specializing to two dimensions in the second line below:
\beqa
\label{gwoperator0}
D_0(p,q) &=& 1 - {1\over \sqrt{X_0 X_0^\dagger}} \;X_0 = \delta_{p,q} \; d_0(p)~,\nonumber \\
d_0(p) &=&  \left( \begin{array}{cc} a(p) & i d(p) + b(p) \cr i d(p) - b(p) & a(p)\end{array}\right)~,
\eeqa
where, consistent with the notation of \cite{Giedt:2007qg}, we have defined the function of momenta:
\beqa
\label{abcw}
a(p) &\equiv & 1 - {1 - 2 s(p_1)^2 - 2 s(p_2)^2 \over w(p) }~, \nonumber\\
b(p) &\equiv & {s(2 p_2) \over w(p)}~,\\
d(p) &\equiv & {s(2 p_1) \over w(p)}~,\nonumber \\
w(p) &\equiv& \sqrt{1 + 8 s(p_1)^2 s(p_2)^2} ~. 
\eeqa
The linear term in (\ref{gwoperator})  is given by:
\beqa
\label{gwoperator1}
D_1(p,q) ={1 \over w(p)+w(q)}\left( - X_1(p,q) + {x_0(p) \over w(p)} \; X_1^\dagger(p,q) \; {x_0(q) \over w(q)} \right)~,
\eeqa
where $X_1^\dagger(p,q) = \gamma_5 X_1(p,q) \gamma_5$, and $w, x_0, X_1$ are as defined above. The second-order term in (\ref{gwoperator}) is then found to be:
\beqa
\label{gwoperator2}
D_2(p,q) &=&{1 \over w(p)+w(q)}\left( - X_2(p,q) + {x_0(p) \over w(p)}\; X_2^\dagger(p,q) \; {x_0(q) \over w(q)} \right) \nonumber \\
&&+ \;  \sum_{\{k_\mu = 1\}}^N \; {1 \over [w(p)+w(q)] [w(p)+w(k)] [w(q)+w(k)]} \times \\
&&~ \left\{ -\; {w(p)+w(q)+w(k) \over w(p)w(q)w(k)} \; x_0(p)\; X_1^\dagger(p,k)\; x_0(k) \; X_1^\dagger(k,q)\; x_0(q)  \right. \nonumber \\
&& \left. + \;
x_0(p) \; X_1^\dagger(p,k) \;  X_1(k,q) + X_1(p,k)\; X_1^\dagger(k,q) \; x_0(q) + X_1(p,k) \; x_0^\dagger(k) \; X_1(k,q) \right\} .\nonumber
\eeqa
Both the linear and quadratic terms in the expansion are consistent with the ones found in \cite{Ishibashi:1999ik}, the precise map being $D_{1,2}^{{\rm this \; paper}} =- \left(D_{1,2}^{{\rm ref.}  \cite{Ishibashi:1999ik}}\right)^\dagger\bigg \vert_{x_0,X_{1,2} \leftrightarrow x_0^\dagger, X_{1,2}^\dagger}$, using our expressions for $x_0, X_1, X_2$.
\section{The chiral eigenvector basis}
\label{eigenvectorbasis}

To   compute the mirror polarization operator, we need explicit expressions for the expansion (\ref{evs1}) in terms of $\hat\gamma_5$ eigenvectors for $A=0$. These were worked out in the Appendix of \cite{Giedt:2007qg} and we give the $A=0$ expansions (\ref{evs1}) here:
\beqa
\label{zeroAexpansions}
 \chi_+(x) &=& \sum_{k} \beta_+^k v_k(x) ~~, ~~ \bar\chi_+(x) = \sum_{k} \bar\beta_+^k u_k^\dagger(x) \\
 \psi_-(x) &=& \sum_{k} \alpha_-^k t_k(x) ~~, ~~ \bar\psi_-(x) = \sum_{k} \bar\alpha_+^k w_k^\dagger(x) ~ \nonumber ,
\eeqa
where the sum is over momenta, $k_1, k_2 = 1 \ldots N$, which label the independent eigenvectors for a trivial gauge background. 

The  $\gamma_5$ ($\hat\gamma_5$) eigenvectors $t,v$ ($w,u$) are orthonormal  (we note that the phase factor in $t_k(x)$ is   a matter of pure convenience and is included for agreement with \cite{Giedt:2007qg}) and are:
\beqa
\label{zeroAeigenvectors}
v_k (x) &=& {1\over N} \; \omega_N^{k \cdot x} \;  \left(\begin{array}{c}1\cr 0\end{array} \right) \nonumber ~,\\
u_k^\dagger (x) &=& {1\over N} \; \omega_N^{- k \cdot x}\;  \left(0 \;\: 1 \right) V_k~,  \\
t_k (x) &=& {1\over N} \; \omega_N^{k \cdot x} \;  e^{- i \varphi_k \sigma_3} \left(\begin{array}{c}0\cr 1\end{array} \right) \nonumber ~,\\
w_k^\dagger (x) &=&  {1\over N} \; \omega_N^{- k \cdot x}\;  \left(1 \;\: 0 \right) U_k ~.  \nonumber~
\eeqa
Here we use the unitary matrices $U_k$ and $V_k$:
\begin{equation}
\label{unitaryprojectors}
U_{k} = {1\over 2} \;   \left( \begin{array}{cc}  2 - \lambda_{{k}} &\lambda_{{k}} 
e^{- i \varphi_{{k}}} \cr - \lambda_{{k}}^* e^{i \varphi_{{k}} }& 2 - \lambda_{{k}}^* \end{array} \right)        
 ~, ~~ V_{{k}}   = {1\over 2} \;   \left( \begin{array}{cc}  (2 - \lambda_{{k}}^*) e^{i \varphi_{{k}} }&- \lambda_{{k}}^*
 \cr  \lambda_{{k}} & (2 - \lambda_{{k}}) e^{- i \varphi_{{k}}}\end{array} \right)        ~,
\end{equation}
  defined in terms of the positive imaginary part eigenvalues of the GW operator (the functions $a, b,d$ are as in (\ref{abcw})):
\begin{equation}
\label{eigenvaluesGW}
\lambda_{{k}} = a(k) + i \; \sqrt{b(k)^2 + d(k)^2}~,
\end{equation}
and the phase factor:
\begin{equation}
\label{phi}
e^{i \varphi_{{k}}} \equiv \left\{ \begin{array}{cc}  {i\; b(k) + d(k) \over \sqrt {b(k)^2 + d(k)^2 } } ,& ~(k_1, k_2) \ne (N,N), (N/2,N/2), (N/2,N), (N,N/2).\cr
1,& ~(k_1, k_2) = (N,N), (N/2,N/2), (N/2,N), (N,N/2)~. \end{array} \right.
\end{equation}
Using these definitions, it is straightforward to verify that the $w, u$ and $t, v$ bases obey $\sum_x (w^\dagger_k (x) \cdot w_p(x)) = \delta_{kp}$, $\sum_x (w^\dagger_k (x) \cdot u_p(x)) = 0$, etc.; proving this only requires (in the $w, u$ case) use of the GW relation $\lambda_k + \lambda^*_k = |\lambda_k|^2$.

\section{Perturbative derivation of the anomaly}
\label{anomalyderivation}

Here, we derive eqn.~(\ref{tl1}).
Now, we know that---see eqn.~(\ref{z13}), as well  as (\ref{piprime5}, \ref{piprime6}): 
\beq
\label{z15}
\sum\limits_\mu \nabla^*_{\mu x} \Pi^{ \prime \; \pm}_{\mu \nu}(x,y) = \mp {i \over 2} \; {\delta {\rm tr} \; \hat{\gamma}_{5}(x,x)[A] \over \delta A_\nu (y)}\bigg\vert_{A = 0}~.
\eeq
For the purpose of comparing with finite-volume simulations of the mirror sector, 
we would like  to have a finite volume expression for the r.h.s. of (\ref{z15}). 

To begin, we first 
define the Fourier transform of any free (i.e. translationally invariant) polarization operator $\Pi_{\mu  \vec{x};    \nu \vec{y} }$  similar to the usual continuum definition:
\beq
\label{FT}
\tilde\Pi_{\mu \nu}(k) \equiv \sum\limits_{{x}} \; \omega_N^{ - k \cdot x} \;  \Pi_{ \mu \nu}( x, 0)~.
\eeq
We then also define the Fourier transform of its divergence:
\beq
\label{divergences}
d_{\nu}(k) \equiv \sum_y \omega_N^{- k \cdot y} \nabla_{\mu x }^*\; \Pi_{\mu  \nu}( x,y)\big\vert_{x=0} = \sum_\mu (1 - \omega_N^{k_\mu}) \tilde{\Pi}_{\mu \nu} (-k)~. \eeq
Finally, we similarly define the  Fourier transform of  the r.h.s. of (\ref{z15}):
\beq
\label{trgamma5ft}
t_\nu(k) \equiv {i \over 2} \;  \sum_y  \omega_N^{- k \cdot y} \; {\delta {\rm tr} \; \hat{\gamma}_5 (0,0)[A] \over \delta A_\nu (y)}\bigg\vert_{A = 0}~,
\eeq
allowing us to recast (\ref{z15}) in the form $d_\nu^\pm(k) = \mp t_\nu(k)$. 

Now we can compute the function $t_\nu(k)$ explicitly, using (\ref{gwoperator}) and (\ref{gwoperator1}) for $D_1$, and find:
\beqa
\label{tnuk}
t_\nu(k) &=&\omega_N^{k_\nu\over 2} \;  {1\over N^2}\sum_{\{p_\mu=1\}}^N {f(p) f(p-k) \over w(p) w(p-k) [w(p) + w(p-k)]} \times \\
&& \left\{  s(2 p_\nu - kŒ_\nu) \; \epsilon^{\alpha\beta}\xi_\alpha(p) \xi_\beta(p-k) - c(2 p_\nu - k_\nu)\;  \epsilon^{\nu\mu} \; \left[ \xi_\mu(p) - \xi_\mu(p-k)\right] \right\}~,
\eeqa
where $\epsilon^{12} = 1$ and we have defined:
\beqa
f(p) &=& 1-2s(p_1)^2 - 2 s(p_2)^2 \\
\xi_\mu(p) &=& {s(2 p_\mu) \over f(p)} ~. \nonumber
\eeqa
It is easy to check that in the small-$k$ limit, eqn. (\ref{tnuk}) reduces to:
\beqa
\label{tnuksmall}
t_\nu(k) &\simeq& \epsilon_{\nu\mu} {2 \pi k_\mu \over N} {1\over 2 N^2}  \sum_{\{p_\alpha =1\}}^N {c(2p_1) + c(2 p_2) - c(2 p_1) c(2 p_2) \over w(p)^3 } \bigg\vert_{N \rightarrow \infty} + {\cal{O}}(k^2) ~\\
&=&{1 \over 2 \pi} \; \epsilon_{\nu\mu} k_\mu^{cont.} ~,\nonumber
\eeqa
where $k_\mu^{cont.} = {2 \pi k_\mu\over N}$ is the continuum momentum; we note that this establishes the coefficient in (\ref{tl1}).


\begin{thebibliography}{99}

  %\cite{Raby:1979my}
\bibitem{Raby:1979my}
 S.~Raby, S.~Dimopoulos,   and L.~Susskind,
  ``Tumbling gauge theories,'' Nucl.\ Phys.\  B {\bf 169}, 373 (1980).
  %%CITATION = NUPHA,B169,373;%%
  
  %\cite{Shifman:1999mv}
\bibitem{Shifman:1999mv}
  M.~A.~Shifman and A.~I.~Vainshtein,
 ``Instantons versus supersymmetry: Fifteen years later,''
  arXiv:hep-th/9902018.
  %%CITATION = HEP-TH/9902018;%%
  
  %\cite{Intriligator:2007cp}
\bibitem{Intriligator:2007cp}
  K.~A.~Intriligator and N.~Seiberg,
 ``Lectures on supersymmetric gauge theories and electric-magnetic  duality,''  Nucl.\ Phys.\ Proc.\ Suppl.\  {\bf 45BC}, 1 (1996)
  [arXiv:hep-th/9509066].
  %%CITATION = NUPHZ,45BC,1;%%
     
     
     %\cite{Shifman:2008cx}
\bibitem{Shifman:2008cx}
  M.~Shifman and M.~\" Unsal,
``On Yang-Mills theories with chiral matter at strong coupling,''
  arXiv:0808.2485 [hep-th].
  %%CITATION = ARXIV:0808.2485;%%

\bibitem{Golterman:2000hr}%Luscher:2000hn
  M.~Golterman,
  ``Lattice chiral gauge theories,''
  Nucl.\ Phys.\ Proc.\ Suppl.\  {\bf 94}, 189 (2001)
  [arXiv:hep-lat/0011027].
  %%CITATION = HEP-LAT 0011027;%%


\bibitem{Luscher:2000hn}
  M.~Luscher,
  ``Chiral gauge theories revisited,''
  arXiv:hep-th/0102028.
  %%CITATION = HEP-TH 0102028;%%
  
  %\cite{Neuberger:2001nb}
\bibitem{Neuberger:2001nb}
  H.~Neuberger,
  ``Exact chiral symmetry on the lattice,''
  Ann.\ Rev.\ Nucl.\ Part.\ Sci.\  {\bf 51}, 23 (2001)
  [arXiv:hep-lat/0101006].
  %%CITATION = ARNUA,51,23;%%



  \bibitem{Eichten:1985ft}
  E.~Eichten and J.~Preskill,
  ``Chiral gauge theories on the lattice,''
  Nucl.\ Phys.\ B {\bf 268}, 179 (1986).
  %%CITATION = NUPHA,B268,179;%%

\bibitem{Smit1}
J.~Smit, ``Fermions on a Lattice," Lectures given at the 25th Cracow International School of Theoretical Physics, Zakopane, Poland, Jun 2-14, 1985, Acta Phys. Polon. B{\bf 17}:531,1986.

\bibitem{Fradkin:1978dv}
  E.~H.~Fradkin and S.~H.~Shenker,
  ``Phase diagrams of lattice gauge theories with Higgs fields,''
  Phys.\ Rev.\ D {\bf 19}, 3682 (1979).
  %%CITATION = PHRVA,D19,3682;%%

\bibitem{Forster:1980dg}
  D.~Foerster, H.~B.~Nielsen and M.~Ninomiya,
  ``Dynamical stability of local gauge symmetry: creation of light from
  chaos,''
  Phys.\ Lett.\ B {\bf 94}, 135 (1980).
  %%CITATION = PHLTA,B94,135;%%
  
%\cite{Lang:1981qg}\cite{Poppitz:2008au}
\bibitem{Lang:1981qg}
  C.~B.~Lang, C.~Rebbi and M.~Virasoro,
 ``The Phase Structure Of A Nonabelian Gauge Higgs Field System,''
  Phys.\ Lett.\  B {\bf 104}, 294 (1981).
  %%CITATION = PHLTA,B104,294;%%
  

%Eichten:1985ft\cite{Hasenfratz:1988vc, Stephanov:1990pc}
\bibitem{Hasenfratz:1988vc}
  A.~Hasenfratz and T.~Neuhaus,
  ``Nonperturbative study of the strongly coupled scalar fermion model,''
  Phys.\ Lett.\ B {\bf 220}, 435 (1989).
  %%CITATION = PHLTA,B220,435;%%



%\cite{Stephanov:1990pc}  
  \bibitem{Stephanov:1990pc}
  M.~A.~Stephanov and M.~M.~Tsypin,
``The phase structure of the $U(1)$ Higgs-fermion lattice theory,''
  Phys.\ Lett.\ B {\bf 242}, 432 (1990).
  %%CITATION = PHLTA,B242,432;%%
  
    %\cite{Poppitz:2008au}
\bibitem{Poppitz:2008au}
  E.~Poppitz and Y.~Shang,
 ``'Light from chaos' in two dimensions,''
  Int.\ J.\ Mod.\ Phys.\  A {\bf 23}, 4545 (2008)
  [arXiv:0801.0587 [hep-lat]].
  %%CITATION = IMPAE,A23,4545;%%

  
\bibitem{Golterman:1990zu}
  M.~Golterman and D.~N.~Petcher,
 ``The 1/d expansion for lattice theories of chiral fermions with Yukawa
  couplings,''
  Nucl.\ Phys.\ B {\bf 359}, 91 (1991).
  %%CITATION = NUPHA,B359,91;%%


%Golterman:1990zu  Golterman:1992yh
\bibitem{Golterman:1991re}
  M.~Golterman, D.~N.~Petcher and J.~Smit,
  ``Fermion interactions in models with strong Wilson-Yukawa couplings,''
  Nucl.\ Phys.\ B {\bf 370}, 51 (1992); 
  %%CITATION = NUPHA,B370,51;%%%\cite{Bock:1992gp}; 
  
  
  \bibitem{Smit2} W.~Bock, A.~K.~De, and J.~Smit, ``Fermion masses at strong Wilson-Yukawa coupling in the symmetric phase," Nucl. Phys. B{\bf 388}, 243 (1992)
  
  \bibitem{Smit1993}
  W.~Bock, A.~K.~De, E.~Focht and J.~Smit,
  ``Fermion Higgs model with strong Wilson-Yukawa coupling in two dimensions,''
  Nucl.\ Phys.\ B {\bf 401}, 481 (1993)
  [arXiv:hep-lat/9210022].
  %%CITATION = HEP-LAT 9210022;%%

  
\bibitem{Golterman:1993th}
M.~Golterman, K.~Jansen, D.~N.~Petcher and J.~C.~Vink,
``Investigation of the domain wall fermion approach to chiral gauge theories
on the lattice,''
Phys.\ Rev.\ D {\bf 49}, 1606 (1994)
[arXiv:hep-lat/9309015].
%%CITATION = HEP-LAT 9309015;%%

\bibitem{Golterman:1994at}
M.~Golterman and Y.~Shamir,
``Domain wall fermions in a waveguide: the phase diagram at large Yukawa
coupling,''
Phys.\ Rev.\ D {\bf 51}, 3026 (1995)
[arXiv:hep-lat/9409013].
%%CITATION = HEP-LAT 9409013;%%

%\cite{Golterman:1992yh}
\bibitem{Golterman:1992yh}
  M.~Golterman, D.~N.~Petcher and E.~Rivas,
  ``Absence of chiral fermions in the Eichten-Preskill model,''
  Nucl.\ Phys.\ B {\bf 395}, 596 (1993)
  [arXiv:hep-lat/9206010].
  %%CITATION = HEP-LAT 9206010;%%

%\cite{Bhattacharya:2006dc}Creutz:1996xc
\bibitem{Bhattacharya:2006dc}
  T.~Bhattacharya, M.~R.~Martin and E.~Poppitz,
  ``Chiral lattice gauge theories from warped domain walls and  Ginsparg-Wilson
  fermions,''
  Phys.\ Rev.\ D {\bf 74}, 085028 (2006)
  [arXiv:hep-lat/0605003].
  %%CITATION = HEP-LAT 0605003;%

\bibitem{Creutz:1996xc}
  M.~Creutz, M.~Tytgat, C.~Rebbi and S.~S.~Xue,
  ``Lattice formulation of the standard model,''
  Phys.\ Lett.\  B {\bf 402}, 341 (1997)
  [arXiv:hep-lat/9612017].
  %%CITATION = PHLTA,B402,341;%%



\bibitem{Ginsparg:1981bj}%\cite{Neuberger:1997fp}
  P.~H.~Ginsparg and K.~G.~Wilson,
  ``A remnant of chiral symmetry on the lattice,''
  Phys.\ Rev.\ D {\bf 25}, 2649 (1982).
  %%CITATION = PHRVA,D25,2649;%%

\bibitem{Kaplan:1992bt}
  D.~B.~Kaplan,
  ``A method for simulating chiral fermions on the lattice,''
  Phys.\ Lett.\ B {\bf 288}, 342 (1992)
  [arXiv:hep-lat/9206013];
    %%CITATION = HEP-LAT 9206013;%
 
\bibitem{Kaplan:1992sg} 
D.~B.~Kaplan,
``Chiral fermions on the lattice,''
Nucl.\ Phys.\ Proc.\ Suppl.\  {\bf 30}, 597 (1993).
%%CITATION = NUPHZ,30,597;%%

\bibitem{Narayanan:1993ss}
  R.~Narayanan and H.~Neuberger,
  ``Chiral fermions on the lattice,''
  Phys.\ Rev.\ Lett.\  {\bf 71}, 3251 (1993)
  [arXiv:hep-lat/9308011]; 
  %%CITATION = HEP-LAT 9308011;%%
  
  \bibitem{NN2}
  R.~Narayanan and H.~Neuberger,
  ``A construction of lattice chiral gauge theories,''
  Nucl.\ Phys.\ B {\bf 443}, 305 (1995)
  [arXiv:hep-th/9411108].
  
%\cite{Neuberger:1997fp}
\bibitem{Neuberger:1997fp}
  H.~Neuberger,
  ``Exactly massless quarks on the lattice,''
  Phys.\ Lett.\ B {\bf 417}, 141 (1998)
  [arXiv:hep-lat/9707022].
  %%CITATION = HEP-LAT 9707022;%%
  
  
\bibitem{Hasenfratz:1998ri}
  P.~Hasenfratz, V.~Laliena and F.~Niedermayer,
  ``The index theorem in QCD with a finite cut-off,''
  Phys.\ Lett.\ B {\bf 427}, 125 (1998)
  [arXiv:hep-lat/9801021].
  %%CITATION = HEP-LAT 9801021;%%
   

\bibitem{Fujikawa:1998if}%Adams:2001jd
  D.~H.~Adams,
  ``Axial anomaly and topological charge in lattice gauge theory with
  overlap-Dirac operator,''
  Annals Phys.\  {\bf 296}, 131 (2002)
  [arXiv:hep-lat/9812003];
  %%CITATION = APNYA,296,131;%%   
  
  \bibitem{F2}
  K.~Fujikawa,
  ``A continuum limit of the chiral Jacobian in lattice gauge theory,''
  Nucl.\ Phys.\ B {\bf 546}, 480 (1999)
  [arXiv:hep-th/9811235].
  %%CITATION = HEP-TH 9811235;%%
  

\bibitem{Adams:2000yi}
  D.~H.~Adams,
  ``Index of a family of lattice Dirac operators and its relation to the
  nonAbelian anomaly on the lattice,''
  Phys.\ Rev.\ Lett.\  {\bf 86}, 200 (2001)
  [arXiv:hep-lat/9910036];  ``Global obstructions to gauge invariance in chiral gauge theory on the
  lattice,''
  Nucl.\ Phys.\  B {\bf 589}, 633 (2000)
  [arXiv:hep-lat/0004015].
  %%CITATION = NUPHA,B589,633;%%
  

\bibitem{Adams:2001jd}
  D.~H.~Adams,
  ``Families index theory for overlap lattice Dirac operator. I,''
  Nucl.\ Phys.\  B {\bf 624}, 469 (2002)
  [arXiv:hep-lat/0109019]; 
  ``Gauge fixing, families index theory, and topological features of the  space
  of lattice gauge fields,''
  Nucl.\ Phys.\  B {\bf 640}, 435 (2002)
  [arXiv:hep-lat/0203014].
  %%CITATION = NUPHA,B640,435;%%


  %
\bibitem{Hernandez:1998et}%\cite{Neuberger:1999pz}
  P.~Hernandez, K.~Jansen and M.~Luscher,
  ``Locality properties of Neuberger's lattice Dirac operator,''
  Nucl.\ Phys.\ B {\bf 552}, 363 (1999)
  [arXiv:hep-lat/9808010].
  %%CITATION = HEP-LAT 9808010;%%

%\cite{Neuberger:1999pz}
\bibitem{Neuberger:1999pz}
  H.~Neuberger,
  ``Bounds on the Wilson Dirac operator,''
  Phys.\ Rev.\ D {\bf 61}, 085015 (2000)
  [arXiv:hep-lat/9911004].
  %%CITATION = HEP-LAT 9911004;%%




%\cite{Giedt:2007qg}
\bibitem{Giedt:2007qg}
  J.~Giedt and E.~Poppitz,
  ``Chiral lattice gauge theories and the strong coupling dynamics of a
  Yukawa-Higgs model with Ginsparg-Wilson fermions,''
  JHEP {\bf 0710}, 076 (2007)
  [arXiv:hep-lat/0701004].
  %%CITATION = JHEPA,0710,076;%%
  
%\cite{Jackiw:1984zi}\cite{Halliday:1985tg}
\bibitem{Jackiw:1984zi}
  R.~Jackiw and R.~Rajaraman,
  ``Vector meson mass generation through chiral anomalies,''
  Phys.\ Rev.\ Lett.\  {\bf 54}, 1219 (1985)
  [Erratum-ibid.\  {\bf 54}, 2060 (1985)].
  %%CITATION = PRLTA,54,1219;%%

%\cite{Halliday:1985tg}
\bibitem{Halliday:1985tg}
  I.~G.~Halliday, E.~Rabinovici, A.~Schwimmer and M.~S.~Chanowitz,
``Quantization of anomalous two-dimensional models,''
  Nucl.\ Phys.\ B {\bf 268}, 413 (1986).
  %%CITATION = NUPHA,B268,413;%%


   %\cite{Poppitz:2007tu}
\bibitem{Poppitz:2007tu}
  E.~Poppitz and Y.~Shang,
  ``Lattice chirality and the decoupling of mirror fermions,''
  JHEP {\bf 0708}, 081 (2007)
  [arXiv:0706.1043 [hep-th]].
  %%CITATION = JHEPA,0708,081;%%
  
  \bibitem{Luscher:1998kn}
  M.~Luscher,
  ``Topology and the axial anomaly in abelian lattice gauge theories,''
  Nucl.\ Phys.\  B {\bf 538}, 515 (1999)
  [arXiv:hep-lat/9808021].
  %%CITATION = NUPHA,B538,515;%%



  
    %\cite{Neuberger:1998xn}Luscher:1998pq
\bibitem{Neuberger:1998xn}
  H.~Neuberger,
  ``Geometrical aspects of chiral anomalies in the overlap,''
  Phys.\ Rev.\  D {\bf 59}, 085006 (1999)
  [arXiv:hep-lat/9802033].
  %%CITATION = PHRVA,D59,085006;%%

\bibitem{Luscher:1998pq}
  M.~Luscher,
  ``Exact chiral symmetry on the lattice and the Ginsparg-Wilson relation,''
  Phys.\ Lett.\ B {\bf 428}, 342 (1998)
  [arXiv:hep-lat/9802011].
  %%CITATION = HEP-LAT 9802011;%%

\bibitem{Lu2}
   M.~Luscher,
   ``Lattice regularization of chiral gauge theories to all orders of
  perturbation theory,''
  JHEP {\bf 0006}, 028 (2000)
  [arXiv:hep-lat/0006014]. 
  
  %\cite{Luscher:1998du}
 \bibitem{Luscher:1998du}
  M.~L\" uscher,
  ``Abelian chiral gauge theories on the lattice with exact gauge
  invariance,''
  Nucl.\ Phys.\  B {\bf 549}, 295 (1999)
  [arXiv:hep-lat/9811032].
  
  \bibitem{GS}
  M.~Golterman and Y.~Shamir, private communication (2007).
  
  %\cite{Gerhold:2007yb,Gerhold:2007gx}
\bibitem{Gerhold:2007yb}
  P.~Gerhold and K.~Jansen,
  ``The phase structure of a chirally invariant lattice Higgs-Yukawa model for
  small and for large values of the Yukawa coupling constant,''
  JHEP {\bf 0709}, 041 (2007)
  [arXiv:0705.2539 [hep-lat]].
  %%CITATION = JHEPA,0709,041;%%
  
  %\cite{Gerhold:2007gx}
\bibitem{Gerhold:2007gx}
  P.~Gerhold and K.~Jansen,
  ``The phase structure of a chirally invariant lattice Higgs-Yukawa model -
  numerical simulations,''
  JHEP {\bf 0710}, 001 (2007)
  [arXiv:0707.3849 [hep-lat]].
  %%CITATION = JHEPA,0710,001;%%  
  
      %\cite{Frishman:1980dq}Coleman:1982yg
\bibitem{Frishman:1980dq}
  Y.~Frishman, A.~Schwimmer, T.~Banks and S.~Yankielowicz,
  ``The axial anomaly and the bound state spectrum in confining theories,''
  Nucl.\ Phys.\  B {\bf 177}, 157 (1981).
  %%CITATION = NUPHA,B177,157;%%

%\cite{Coleman:1982yg}
\bibitem{Coleman:1982yg}
  S.~R.~Coleman and B.~Grossman,
 ``'t Hooft's consistency condition as a consequence of analyticity and
   unitarity,''
  Nucl.\ Phys.\  B {\bf 203}, 205 (1982).
  %%CITATION = NUPHA,B203,205;%%
      
 

  
  %\cite{Rebbi:1986ra}
\bibitem{Rebbi:1986ra}
  C.~Rebbi,
 ``Chiral invariant regularization of fermions on the lattice,''
  Phys.\ Lett.\  B {\bf 186}, 200 (1987).
  %%CITATION = PHLTA,B186,200;%%
  
  %\cite{Bodwin:1987gi}
\bibitem{Bodwin:1987gi}
  G.~T.~Bodwin and E.~V.~Kovacs,
   ``Perturbative analysis of Rebbi's lattice fermion proposal,''
  Phys.\ Lett.\  B {\bf 193}, 283 (1987).
  %%CITATION = PHLTA,B193,283;%%
  
  %\cite{Pelissetto:1987ad}
\bibitem{Pelissetto:1987ad}
  A.~Pelissetto,
  ``Lattice nonlocal chiral fermions,''
  Annals Phys.\  {\bf 182}, 177 (1988).
  %%CITATION = APNYA,182,177;%%
%\cite{Ishibashi:1999ik}
\bibitem{Ishibashi:1999ik}
  M.~Ishibashi, Y.~Kikukawa, T.~Noguchi and A.~Yamada,
  ``One-loop analyses of lattice QCD with the overlap Dirac operator,''
  Nucl.\ Phys.\  B {\bf 576}, 501 (2000)
  [arXiv:hep-lat/9911037].
  %%CITATION = NUPHA,B576,501;%%


%\cite{Kikukawa:1999sy}
\bibitem{Kikukawa:1999sy}
  Y.~Kikukawa and T.~Noguchi,
  ``Low energy effective action of domain-wall fermion and the  Ginsparg-Wilson
   relation,''
  arXiv:hep-lat/9902022.
  %%CITATION = HEP-LAT/9902022;%%

%\cite{Fosco:2007ry}
\bibitem{Fosco:2007ry}
  C.~D.~Fosco, G.~Torroba and H.~Neuberger,
  ``A simple derivation of the overlap Dirac operator,''
  Phys.\ Lett.\  B {\bf 650}, 428 (2007)
  [arXiv:0704.2433 [hep-lat]].
  %%CITATION = PHLTA,B650,428;%%

  
%\cite{Chiu:1998aa}
\bibitem{Chiu:1998aa}
  T.~W.~Chiu, C.~W.~Wang and S.~V.~Zenkin,
  ``Chiral structure of the solutions of the Ginsparg-Wilson relation,''
  Phys.\ Lett.\  B {\bf 438}, 321 (1998)
  [arXiv:hep-lat/9806031].
  %%CITATION = PHLTA,B438,321;%%

\end{thebibliography}
\end{document}